\documentclass[nofootinbib,prd,twocolumn,showpacs,showkeys,preprintnumbers]{revtex4-1}
\usepackage{hyperref,amssymb,amsmath,mathrsfs,bm,graphicx}
\usepackage[dvipsnames]{xcolor}
\usepackage{soul}

\begin{document}

\title {Hyperbolic polytrope}

\author{M. Carrasco-H }
\email{mcarrasco@yachaytech.edu.ec}
\affiliation{School of Physical Sciences and Nanotechnology, Yachay Tech University,  Urcuqu\'i, 100119, Ecuador\\}

\author{E. Contreras }
\email{ernesto.contreras@ua.es }
\affiliation{Departamento de F\'{\i}sica Aplicada, Universidad de Alicante, Campus de San Vicente del Raspeig, E-03690 Alicante, Spain.\\}

\author{E. Fuenmayor}
\email{ernesto.fuenmayor@ciens.ucv.ve}
\affiliation{Centro de F\'isica Te\'orica y Computacional,\\ Escuela de F\'isica, Facultad de Ciencias, Universidad Central de Venezuela, Caracas 1050, Venezuela}

\author{P. Le\'on}
\email{pleon@pitp.ca}
\affiliation{Perimeter Institute for Theoretical Physics, Waterloo, ON N2L 2Y5, Canada\\}

\begin{abstract}
In this work, we study self-gravitating objects that obey a polytropic equation of state in hyperbolic symmetry. Specifically, we describe in detail the steps to derive the Lane-Emden equation from the structure equations of the system. To integrate the equations numerically, we propose the Cosenza-Herrera-Esculpi-Witten anisotropy and study the cases
$\gamma \ne 1$ and $\gamma = 1$ in the parameter space of the models. We find that the matter sector exhibits the usual and expected behavior for certain values in this parameter space: energy density (in absolute value) and radial pressure are decreasing functions and vanish at the surface, while the mass function is increasing toward the surface. We find that the  anisotropy of the system is positive and decreasing, consistent with the behavior of the radial pressure, which reaches a local minimum at the surface (i.e., the pressure gradient is zero at the surface). We also study the compactness of the dense objects as a function of the polytropic index and obtain that it has an upper bound given by the maximum value it reaches for a certain $n$. Some extensions of the work and future proposals are discussed.
\end{abstract}


\maketitle

\section{Introduction}

Static solutions in general relativity are of great importance for several reasons. First, their mathematical simplicity makes them easier to analyze since temporal invariance reduces the complexity of Einstein field equations, allowing exact solutions to be obtained in certain cases. Additionally, many static solutions serve as models for astrophysical objects. For example, the exterior Schwarzschild solution describes the spacetime around a non-rotating spherical mass, such as a planet or a star, and is fundamental for understanding phenomena like planetary orbits and the bending of light near massive objects. Static solutions are also essential for describing black holes. Again, the vacuum Schwarzschild solution is the simplest one for a non-charged, non-rotating black hole, and studying these solutions helps us to understand properties such as the event horizon and the singularity of black holes. Furthermore, these solutions allow us to investigate the structure of spacetime in the presence of different configurations of matter and energy, helping us comprehend how mass and energy affect the geometry of spacetime. Often, static solutions serve as starting points for finding more complex solutions. For example, the Kerr solution, which describes a rotating black hole, can be understood as a generalization of the Schwarzschild solution. Finally, predictions based on static solutions have been experimentally verified, providing crucial evidence for the validity of general relativity.\\

According to Birkhoff's theorem, the exterior Schwarzschild solution is the only static, spherically symmetric, and asymptotically flat solution to Einstein equations in vacuum. However, although the metric does not explicitly depend on time, the static nature, in that an observer can move freely throughout the entire spacetime, fails when crossing the event horizon. To be more precise, in physical coordinates, the Schwarzschild metric is given by
\begin{equation}
ds^{2}=-\left(1-\frac{2M}{r}\right)dt^{2}+\frac{dr^{2}}{1-\frac{2M}{r}}+r^{2}(d\theta^{2}+\sin^{2}\theta d\phi^{2})
\end{equation}
where the coordinate time $t$ is adapted to the timelike Killing vector of the metric. It is evident that at $r=2M$ the metric is singular, but this singularity is removable through a change of coordinates since there is no algebraic singularity at the horizon (the physically meaningful components of the Riemann tensor are regular). The region $r=2M$ is a Killing horizon since there is a change in the signature of the metric when crossing it. This change causes the vector to switch from timelike to spacelike, which means any geodesic crossing $r=2M$ will end at the singularity $r=0$ (which is an essential singularity). In this sense, observers crossing the horizon cannot escape or remain static but will inevitably end at $r=0$. For this reason, although observers can be considered static outside the horizon, they cannot be within it, making the spacetime not globally static \cite{Herrera:2018mzq}. Even more, Rosen states \cite{Rosen:1970nvr} that any transformation that maintains the static form of the Schwarzschild metric and removes the singularity at the horizon is necessarily non-static for $r<2M$ and requires that all incoming geodesics end at the essential singularity. In fact, various
coordinate transformations have been found that allow the manifold to extend over the whole region although all of them necessarily are nonstatic.\\

In a recent work \cite{Herrera:2018mzq} the authors propose a different interpretation of the situation within the horizon. The resulting static solution describes spacetime as consisting of a complete four-dimensional manifold on the exterior side and a second complete four-dimensional solution on the interior. The two-dimensional $r-t$ submanifold forms a single continuous topological manifold and the $\theta-\phi$ submanifolds exhibit spherical symmetry on the exterior and hyperbolic symmetry on the interior, namely the line element is expressed as
\begin{equation}\label{vacioH}
ds^{2}=\left(\frac{2M}{r}-1\right)dt^{2}-\frac{dr^{2}}{\frac{2M}{r}-1}-r^{2}(d\theta^{2}+\sinh^{2}\theta d\phi^{2}).
\end{equation}
These two regions intersect only at a single curve, specifically at $r = 2M $ and $\theta = 0$. A geodesic can only traverse from one side of the  $r-t$  manifold to the other at this curve. The horizon is meaningful as such solely for static geodesic observers. This assumption carries profound implications. After a comprehensive study of the geodesic behavior of test particles in this globally static spacetime, it becomes evident that the kinematic and dynamic properties of a test particle inside the horizon are significantly different. Notably, test particles are not destined to move toward the center, and cannot reach it for any finite amount of energy, as demonstrated in \cite{Herrera:2020bfy}. This phenomenon arises from a repulsive force within the horizon that pushes test particles away from the center, potentially related to the quantum vacuum of the gravitational field. Furthermore, particles within the horizon may, in principle, exit that region along the axis $\theta = 0$. Thus, a particle could come 
from $\infty$, cross the horizon, bounce back before reaching the center, and then cross the horizon again outward. However, this is only possible along this axis; all other trajectories remain confined within the horizon. 

An interesting question to consider is, given the intriguing physics that develops inside the horizon of this globally static spacetime, what is the source associated with (\ref{vacioH})? In other words, what is the behavior of hyperbolic-symmetric fluids that support the hyperbolic vacuum? These aspects were analyzed in \cite{Hyperbolic}, where the field equations, Tolman mass, orthogonal decomposition of the Riemann tensor, and some specific solutions were studied, such as solutions with incompressible fluids, solutions with vanishing complexity and solutions with matter sectors obeying a stiff equation of state. Among the conclusions, they found that such fluid distributions can be anisotropic in pressure, with only two principal stresses being unequal and the energy density is necessarily negative. Moreover, the fluid cannot occupy the entire space within the horizon, excluding the central region. This holds true regardless of whether the energy density within the fluid distribution is regular or not. This result implies that the central region must consist of a  cavity  described by a different type of source. Additionally, the fact that the fluid distribution cannot reach the center aligns with the result that no test particle with finite energy can reach the center. In this regard, the whole spacetime is composed of different submanifolds $\mathcal{M}_{cavity} \cup \mathcal{M}_{fluid} \cup \mathcal{M}_{h-vacuum} \cup \mathcal{M}_{s-vacuum}$, where {\it cavity} corresponds to the central cavity, {\it fluid} is the region of the fluid under consideration, {\it h-vacuum} corresponds to the hyperbolic vacuum, and {\it s-vacuum} corresponds to the Schwarzschild vacuum solution. If the submanifold $\mathcal{M}_{cavity}$ is spherically symmetric (as we will discuss later) it only meet $\mathcal{M}_{fluid}$ at $\theta=0$. Clearly, $\mathcal{M}_{h-vacuum}$ and $\mathcal{M}_{s-vacuum}$ only meet at $\theta=0$.\\

Our purpose here is to carry out a complete and detailed study on the physical properties and thermodynamic behavior of the variables of a fluid distribution in the region inner to the horizon, endowed with the hyperbolic symmetry and characterized mainly by a polytropic equation of state. Such a fluid distribution might serve as a source of the hyperbolic vacuum metric within the horizon given by (\ref{vacioH}). We are interested in studying static fluids with hyperbolic symmetry that obey a polytropic equation of state, given the importance of such fluids in relativistic astrophysics within the context of static, spherically symmetric solutions. Polytropic equations of state have a rich and esteemed history in astrophysics (see \cite{cha, 3, 7b} and references therein) and have been widely applied to investigate stellar structures under various conditions, initially focusing on Newtonian fluids within isotropic stellar bodies. However, it is now recognized that pressure isotropy might be an overly restrictive assumption. Pressure anisotropy arises from numerous physical effects, particularly in extremely compact objects (see \cite{14, 1p, 5p, 8p} and references therein). Recent studies have shown that isotropic pressure conditions become unstable due to factors such as dissipation, inhomogeneity in energy density and shear \cite{LHP}. This has led to a renewed interest in examining fluids that do not meet the isotropic pressure criterion and has prompted the expansion of polytrope theory to include anisotropic fluids. Newtonian polytropes are applicable to self-gravitating objects with compactness similar to or less than that of white dwarfs. In such cases, Newtonian gravity offers a satisfactory approximation for the gravitational interactions of these stellar objects. However, for more compact objects, such as neutron stars, super-Chandrasekhar white dwarfs or strange quark stars for example, where relativistic effects become significant, it is necessary to extend polytrope theory into the realm of general relativity. General relativistic polytropes have been extensively investigated (see \cite{4a, 4b, 4c, 5, 6, 7, 11, 12, 13, 2p, 3p, N1, H1, 6p, B1, M1, 7p, LN3, efl, LN2} and references therein), notably, \cite{2p} provides a detailed and comprehensive study on general relativistic polytropes for anisotropic fluids. Also, a very complete analysis with varied applications has been published a few years ago (\cite{efl, LN2}) which includes a generalized equation that has been used for various particular cases \cite{45,46}. Finally, it is worth mentioning that in reference \cite{ILEP} is taken an alternative route to integrate the Lane–Emden equation which consists of considering stellar interiors supported by anisotropic fluids fulfilling the polytropic equation of state in combination with the Minimal Geometric Deformation approach \cite{JO}, which represents a novel approach to get families of relativistic anisotropic polytropes. 

This work is organized as follows. In the next section,  we review the main aspects of fluid distributions in hyperbolic symmetry developed in \cite{Hyperbolic}. Next, in section \ref{politropo}, we formulate the polytropic equations and the corresponding Lane-Emden equation. Then, in section \ref{chew} we consider the Cosenza-Herrera-Esculpi-Witten anisotropy and integrate the Lane-Emden equation for different values of the parameters involved. Section \ref{discussion} is devoted to the discussions of the results and in the last section we conclude the work.


\section{Field equations}

We consider hyperbolically symmetric distributions of static fluid, locally anisotropic and which may be bounded from the exterior by a surface $\Sigma^{e}$ whose
equation is $r = r_{\Sigma^e}$ = constant (which is the outer boundary of $\mathcal{M}_{fluid}$). On the other hand, the fluid distribution cannot fill the central region, in which case we may assume that such a region is represented by another distribution of matter with different symmetry, implying that the fluid distribution is also bounded from the inside by a surface $\Sigma^i$ whose equation is $r = r_{\Sigma^i}=constant$ (which is the inner boundary of $\mathcal{M}_{fluid}$). Here we propose the physical variables and the basic set of equations required to describe this hyperbolically symmetric, static and locally anisotropic matter distribution. The metric, in polar coordinates, is given by
\begin{eqnarray}
ds^2 = e^\nu dt^2 -e^\lambda dr^2 -r^2(d\theta^2 + \sinh^2{\theta}d\phi^2), \label{lem}
\end{eqnarray}
where, due to the assumed symmetry, $\nu(r)$ and $\lambda(r)$ are functions of $r$ only. 

The metric (\ref{lem}) has to satisfy the Einstein field equations
\begin{eqnarray}
R_{\mu \nu} -\frac{1}{2}Rg_{\mu \nu} = 8 \pi T_{\mu \nu}. \label{Eq}
\end{eqnarray}
The most general canonical algebraic decomposition of
the second order symmetric energy-momentum tensor tensor, satisfying our assumptions, is given by
\begin{eqnarray}
T_{\mu \nu} = (\mu + P_\perp) V_\mu V_\nu -P_\perp g_{\mu \nu} + (P_r -P_\perp) K_\mu K_\nu, \label{emt}
\end{eqnarray}
where $\mu$, $P_r$ and $P_\perp$ are the energy density, the radial pressure and the tangential pressure, respectively. Due to the symmetry under consideration, the physical variables can only denpend on $r$. On the other hand, $V_\mu$ are the four velocity components, which in our case is given by
\begin{eqnarray}
V_\mu = e^{\nu/2}\delta_\mu^0,
\end{eqnarray}
and 
\begin{eqnarray}
K_\mu =  -e^{\lambda/2}\delta_\mu^1,
\end{eqnarray}
together with the vectors  
\begin{eqnarray}
L_\mu = -r\delta^2_\mu, \quad S_\mu = -r\sinh{\theta}\delta^3_\mu, 
\end{eqnarray}
we can define a canonical orthonormal tetrad.

Introducing (\ref{lem}) and (\ref{emt}) in the system of field equations (\ref{Eq}), it can be shown that the only non-vanishing components of Einstein equations are given by
\begin{eqnarray}
8\pi \mu &= & - \frac{(e^{-\lambda}+1)}{r^2}+\frac{\lambda'}{r}e^{-\lambda}, \label{ec1} \\
8\pi P_r &= &  \frac{(e^{-\lambda}+1)}{r^2} + \frac{\nu'}{r}e^{-\lambda},  \label{ec2}\\
8\pi P_\perp & = & \frac{e^{-\lambda}}{2}\left(\nu'' + \frac{\nu'^2}{2}-\frac{\lambda'\nu'}{2}+\frac{\nu'}{r}-\frac{\lambda'}{r}\right), \label{ec3}
\end{eqnarray}
where primes indicate derivatives respect to $r$. It is worth stressing the differences between these equations and the corresponding to the spherically symmetric
case (see for example \cite{efl}). From these equations or using the conservation laws $T^{\mu}_{\nu;\mu} = 0$, we can get the generalized Tolman-Oppenheimer-Volkoff hydrostatic equilibrium equation for anisotropic matter
\begin{eqnarray}
P'_r + (P_r + \mu)\frac{\nu'}{2}+\frac{2}{r}\Pi = 0, \label{TOV}
\end{eqnarray}
where $\Pi =-\Delta= P_r -P_\perp$.

Following the results presented in \cite{Hyperbolic}, we shall define the mass function as 
\begin{eqnarray}
m(r) = -\left(\frac{r}{2}\right) R^3_{232} = \frac{r(1+e^{-\lambda})}{2}, \label{m1}
\end{eqnarray}
where the Riemann tensor component $R_{3232}$, has been calculated with (\ref{lem}). Thus, using (\ref{ec1}), we can write
\begin{eqnarray} \label{mp}
m'(r) = -4\pi r^2 \mu. \label{m2}
\end{eqnarray}

Now, using the mean value theorem to evaluate the mass integral the regularity at the center of the distributions implies that the mass function must vanish as $r^3$. However, from  (\ref{m1}) it is clear that this is not our case. Thus, the hyperbolically symmetric fluid cannot fill the space surrounding the center. Thus, as discussed in \cite{Hyperbolic}, there should be cavity either empty, or filled with a fluid not endowed with hyperbolic symmetry. The situation just described is fully consistent with the results obtained in \cite{Herrera:2020bfy} where it was shown that test particles cannot reach the center for any finite value of its energy.

On the other hand, from (\ref{m2}), it is clear that to get a positive mass function we need $\mu\leq 0$. Thus, there is a violation of the week energy condition. Furthermore, we can write the mass function as
\begin{eqnarray}\label{m3}
m = 4\pi \int_{r_{min}}^r |\mu|r^2dr,
\end{eqnarray}
where due to the fact that $\mu$ is negative, we have replaced it by $-|\mu|$ (as we shall do from now on). Next, using (\ref{ec2}) and (\ref{m1}) we obtain
\begin{eqnarray}\label{nuprima}
\frac{\nu '}{2}=\frac{4\pi r^3 P_r - m}{r(2m-r)},
\end{eqnarray}
from which we may write (\ref{TOV}) as
\begin{eqnarray}
P'_r + (P_r-|\mu|)\frac{4\pi r^3 P_r - m}{r(2m-r)}+\frac{2}{r}\Pi = 0. \label{TOV2}
\end{eqnarray}
This represents the hydrostatic equilibrium equation for our fluid. The first term in (\ref{TOV2}) is just the gradient of pressure (usually negative and opposing gravity). The second term describes the gravitational ``force'' and contains two different contributions. The first one is the ``passive gravitational'' mass density $P_{r} - |\mu|$ which we expect to be negative. Also the term $4\pi r^{3} P_{r} - m$ (composed by the self–regenerative pressure effect $4\pi r^{3} P_r$ minus the mass function) that is proportional to the ``active gravitational mass'', and which is negative if $4\pi r^{3} P_r < m$. As a consequence, the whole second term is positive (as expected). However, because of the equivalence principle, we must keep in mind what a negative passive gravitational mass means in the  hydrostatic description compared to their usual roles with respect to the positive energy density case. Finally the third term describes the effect of the pressure anisotropy. 

Just as we have mentioned, we assume that the fluid is bounded by a hypersurface $\Sigma^e$. Moreover, we will suppose that the space for $r >r_{\Sigma^e}$ is described by the line element
\begin{eqnarray}
ds^2 &=&  \left(\frac{2M}{r}-1\right)dt^2 -\left(\frac{2M}{r}-1\right)^{-1}dr^2 \\ &-& r^2(d\theta^2 + \sinh^2{\theta}d\phi^2).
\end{eqnarray}
Thus the matching conditions lead to 
\begin{eqnarray}\label{emc}
e^{\nu_{\Sigma^e}}  &=& \frac{2M}{r_{\Sigma^e}}-1, \quad e^{-\lambda_{\Sigma^e}} = \frac{2M}{r_{\Sigma^e}}-1, \quad P_r(r_{\Sigma^e})=0. \nonumber \\ &&
\end{eqnarray}
Assuming the inner region is filled with a spherically symmetric fluid (that includes the case of an empty cavity) with a line element given by

\begin{eqnarray} \label{ssle}
    ds^2 = e^{\nu_s}dt^2- e^{\lambda_s}dr^2-r^2(d\theta^2+\sin^2{\theta}d\phi^2),
\end{eqnarray}
where $\nu_s$ and $\lambda_s$ are function of $r$ only. 
As we discussed before, in this case $\mathcal{M}_{cavity}$ and $\mathcal{M}_{fluid}$ only meet at $\theta=0$. Let $\Sigma^{c}$ be the surface of the cavity. At $\theta=0$ we have 
\begin{eqnarray} \label{intme}
   && e^{\nu_s}\big|_{r_{\Sigma^c}} = e^{\nu}\big|_{r_{\Sigma^i}}, \; e^{\lambda_s}\big|_{r_{\Sigma^c}} = e^{\lambda}\big|_{r_{\Sigma^i}}, \; P_r^s(r_{\Sigma^c}) = P_r(r_{\Sigma^i}), \nonumber \\ &&
\end{eqnarray}
where $P_r^s$ is the radial pressure computed with (\ref{ssle}).  Now, let us analyse what we believe are  the simplest possibilities for the the cavity around the center:

\begin{enumerate}
    \item An empty flat cavity
   \begin{eqnarray} \label{mac1}
    e^{\nu_s}=e^{\lambda_s}=1.
    \end{eqnarray}
In addition, the condition $m(r_{\Sigma^i})=0$ must be satisfied. From (\ref{m1}), this condition implies that $e^{\lambda}=-1$, which contradicts (\ref{mac1}). Moreover, this will imply that the metric signature for the hyperbolic fluid becomes $(+,+,-,-)$. For these reasons we rule out this case.
    \item An isotropic fluid with constant density

    \begin{eqnarray}\label{si}
    e^{\nu_s} = \left(\Tilde{A}-B\sqrt{1-\frac{r^2}{C^2}}\right)^2, \quad e^{-\lambda_s} = 1-\frac{r^2}{C^2},
    \end{eqnarray}
    where the matching conditions leads to 

    \begin{eqnarray}
        |\Tilde{A}| &=& \frac{e^{\nu(\Sigma^i)/2}}{4}\left|\frac{m(\Sigma^i)(4-2r_{\Sigma^i}\nu'(\Sigma^i))+r_{\Sigma^i}(r_{\Sigma^i}\nu'(\Sigma^i)-4)}{r_{\Sigma^i}-m(\Sigma^i)}\right|, \nonumber \\ && \\
        |B| &=& \frac{e^{\nu(\Sigma^i)/2}r_{\Sigma^i}^{3/2}\sqrt{2m(\Sigma^i)-r_{\Sigma^i}}|\nu'(\Sigma^i)|}{4|r_{\Sigma^i}-m(\Sigma^i)|}, \\
        C^2 &=& \frac{r_{\Sigma}^3}{2(r_{\Sigma^i} - m(\Sigma^i))}.
    \end{eqnarray}
    
\end{enumerate}
The metric (\ref{si}) corresponds to the well-known Schwarzschild interior solution with arbitrary coefficients. What is interesting about this case is that it presents a layered model where an incompressible fluid (Schwarzschild solution), a hyperbolic polytrope, a hyperbolic vacuum, and finally a spherical vacuum coexist. It is noteworthy that this model resembles the original gravastar solution by Mazur and Mottola \cite{Mazur:2001fv}, with the difference that, instead of the Schwarzschild interior solution approaching the de Sitter solution in the limit where the star's radius tends to the Schwarzschild radius, in our case, the Schwarzschild interior solution is maintained but shielded by the hyperbolic polytrope and the hyperbolic vacuum. In conclusion, we can see that our solution can also be considered a type of gravastar. It is important to note that we could also consider any fluid in spherical symmetry. We chose the Schwarzschild interior solution because it is the simplest and perhaps the most interesting case because of the connection it has with previous similar scenarios as we have just discussed.

Finally, it is important to mention that there is another definition to describe the energy content of a fluid, proposed by Tolman \cite{Tolman}. The general expression for our interior hyperbolic solution is given by 
\begin{eqnarray}\label{mT1}
m_T &&= \int^{2\pi}_0 \int_0^\pi \int_{r_{min}}^r \sqrt{-g}(-|\mu| + P_r +2P_\perp)d\tilde{r}d\tilde{\theta}d\tilde{\phi} \nonumber\\
&&=2\pi (\cosh\pi -1)\, \times \nonumber\\ 
&&\qquad\quad\quad\int_{r_{min}}^r e^{(\nu+\lambda)/2} \tilde{r}^2 (-|\mu| + P_r +2P_\perp)d\tilde{r}.\nonumber\\
\end{eqnarray}
Using (\ref{lem}) and the field equations (\ref{ec1})-(\ref{ec3}), the integration of (\ref{mT1}) allows us to express the Tolman mass by
\begin{eqnarray}\label{mT2}
m_T = \frac{\cosh{\pi}-1}{4}e^{(\nu-\lambda)/2}r^2 \nu ',
\end{eqnarray}
or combining (\ref{nuprima}) with (\ref{mT2})
\begin{eqnarray}\label{mT3}
m_T = \frac{\cosh{\pi}-1}{2}e^{(\nu+\lambda)/2}(4\pi  r^3 P_r - m),
\end{eqnarray}
from which becomes evident the interpretation of the Tolman mass as a measure of the active gravitational mass. Notice that, because of the hyperbolic symmetry (provided $4\pi P_r r^3 < m$), this quantity is negative. Thus, this indicates the repulsive character of the gravitational interaction on our models. Even more, let us consider the four-acceleration $a_{\alpha}=V_{\alpha ; \beta} V^{\beta}$, that in our case may be written as $a_{\alpha}=a K_{\alpha}$, where
\begin{eqnarray}\label{a1}
a = \frac{\nu '}{2} e^{- \lambda/2}
\end{eqnarray}
which, using (\ref{mT2}), allows us to write
\begin{eqnarray}\label{a2}
a = \frac{2m_T}{r^2}\frac{e^{- \nu/2}}{\cosh{\pi}-1}.
\end{eqnarray}
Thus the four-acceleration is directed inwardly when $4\pi P_r r^3 < m$. This is a result consistent with the fact that the Tolman mass is negative, since $a_{\mu}$ represents the inertial radial acceleration which is necessary in order to maintain static the frame by canceling the gravitational acceleration exerted on it, so this again reveals the repulsive nature of the gravitational force. 

Following a procedure analogous to that carried out in Ref. \cite{15}, but now done for the hyperbolic symmetry, we may obtain another useful expression for the Tolman mass (see Ref. \cite{Hyperbolic} for details) ,
\begin{eqnarray}\label{mT4}
m_T &=& (m_T)_{\Sigma^{e}}\left(\frac{r}{r_{\Sigma^{e}}}\right)^3 + \frac{\cosh{\pi}-1}{2} r^3 \nonumber\\
&& \quad\times  \int_{r}^{r_{\Sigma^{e}}} \frac{e^{(\nu+\lambda)/2}}{\tilde{r}}\left( \frac{4\pi}{\tilde{r}^3}\int_{r_{min}}^{\tilde{r}}|\mu |' s^3 ds + 8 \pi \Pi \right)d\tilde{r}.\nonumber\\
\end{eqnarray}


\section{Hyperbolically symmetric polytrope}\label{politropo}

In this section, we shall discuss the formulation of hyperbolically symmetric static matter distributions satisfying a polytropic equation of state. Now, as we mentioned before, the systems with this symmetry have negative energy density. Thus, the first step will be to write a proper form for the polytropic equation of state. Indeed, we start from the standard polytropic equation of state for the radial pressure
\begin{equation}\label{pos}
    P_r = K \mu^{\gamma}= K \mu^{1+1/n},
\end{equation}
where the constants $K$, $\gamma$ and $n$ are usually called the polytropic constant, polytropic exponent, and polytropic index, respectively. Once the equation of state (\ref{pos}) is assumed, the whole system is described by the Lane–Emden equation (which essentially constitutes the dimensionless
form of Tolman–Oppenheimer–Volkoff expression for a polytrope) that may be solved for any set of the parameters of the theory. It is evident that a negative energy density will lead to a complex pressure for some values of $n$. This is 
\begin{eqnarray}\label{pos2}
P_r = K (-1)^{1+1/n}|\mu|^{1+1/n}.
\end{eqnarray}
In order to avoid this problem, we must impose the following condition
\begin{eqnarray}
  Im(\tilde{K}) = 0, \quad \tilde{K} = (-1)^{1+1/n} K.  
\end{eqnarray}
Then, we can write (\ref{pos2}) as 
\begin{eqnarray}\label{Pr}
  P_r = \tilde{K} |\mu|^{1+1/n} = \tilde{K} |\mu|^{\gamma}.
\end{eqnarray}

As is well known from the general theory of polytropes \cite{2p,6p} to obtain the solutions of the main Lane-Emden equation there is a bifurcation at the value of the exponent  $ \gamma = 1$. Thus, the cases $ \gamma = 1$ and $ \gamma \ne 1$ have to be considered separately. Moreover, from $\tilde{K}$ definition, it is important to notice that $P_r$ can be negative. This detail will be discussed in the following sections.


\subsection{The case $\gamma \not=1$}

We are interested in matter configurations satisfying, $\mu(r_{\Sigma^i}) = \mu_s(r_{\Sigma^i}):=\mu_0$ and $\mu(r_{\Sigma^e})=0$. Here, $\mu_s$ denote the energy density obtained by using (\ref{ssle}). Then, for the case $\gamma \ne 1$ we can define
\begin{eqnarray}\label{omega1}
\omega^n = \frac{\mu}{\mu_{f}} = \frac{|\mu|}{|\mu_{f}|},
\end{eqnarray}
where $\mu_{f}\not=0$ is the value of the energy density for some $r_{f}\in [r_{\Sigma^i},r_{\Sigma^e})$. Now, following the same idea presented in \cite{1p,2p}, we shall define some dimensionless variables,


\begin{eqnarray}\label{variables}
r &=& A x, \label{variable1}\\
m &=& 4\pi A^3 |\mu_f| \eta,\label{variable2}\\ 
q&=&\frac{P_r^f}{|\mu_f|},\label{variable3}
\end{eqnarray}

\begin{eqnarray}
A^2 = \frac{q(n+1)}{4\pi |\mu_f|}.
\end{eqnarray}

Introducing these variables in eqs. (\ref{m2}) and  (\ref{TOV2}) we can get the corresponding Lane-Emden system of equations

\begin{eqnarray}
\frac{d\omega}{dx} + &&\frac{( q x^3\omega^{n+1}-\eta)( q\omega-1)}{x[2q(n+1)\eta -x]}\nonumber\\
&& \qquad\quad + \frac{2\Pi}{ x P^f_r\omega^n (n+1)} = 0, \label{le1}
\end{eqnarray}

\begin{eqnarray}
\frac{d\eta}{dx} = x^2 \omega^n. \label{le2}
\end{eqnarray}

Now, in order to write the Tolman mass (\ref{mT1}) using the new variables, we need to provide an expression for $\nu$. This can be done by introducing the dimensionless variables in (\ref{TOV}) and rewriting it in following way 

\begin{eqnarray}
\frac{d\nu}{dx} = -\frac{2q(n+1)}{q\omega-1}\left[\frac{d\omega}{dx}+\frac{2\Pi}{x(n+1)\omega^n  P^f_r}\right].
\end{eqnarray}
Here, the integration can be done by taking as reference any of the two hypersurfaces defined as $x = x_{\Sigma^i}$ or $x = x_{\Sigma^e}$. In this work, we will choose the exterior one.  Then, for $q\omega-1 \not = 0$, we can write

\begin{eqnarray}
\nu(x_{\Sigma^e}) - \nu(x) = -2 q(n+1)\int_x^{x_{\Sigma^e}}\frac{d\omega}{q\omega -1}-G. 
\end{eqnarray}
where
\begin{eqnarray} \label{Gn1}
G = \frac{4q}{P^f_r}\int_x^{x_{\Sigma^e}}\frac{\Pi}{\bar{x}\omega^n ( q\omega -1)}d\bar{x}.
\end{eqnarray}
Now, defining the following variables
\begin{eqnarray}
\eta_T &=& \frac{m_T}{2\pi |\mu_f| A^3 (\cosh{\pi}-1)}, \label{atm}\\
y &=& \frac{M}{r_{\Sigma^e}}, \label{supr}\\ 
z &=& \frac{x}{x_{\Sigma^e}}, \label{nrad}
\end{eqnarray}
and using the matching conditions, we can write
\begin{eqnarray} \label{TMn1}
\eta_T &=& \frac{\sqrt{2y-1}}{|q\omega-1|^{n+1}}e^{G/2}\bigg[\frac{x_{\Sigma^e}z}{2q(n+1)\eta-x_{\Sigma^e}z}\bigg]^{1/2} \nonumber \\ &\times& ( qx^3_{\Sigma^e}z^3\omega^{n+1}-\eta),
\end{eqnarray}

Notice, that the compactenss, $y$, may also been written as
\begin{eqnarray}
y = q(n+1)\eta(x_{\Sigma^e})/x_{\Sigma^e}.
\end{eqnarray}

Then,
\begin{eqnarray}
    e^{\nu(x)}=\frac{(2y-1)e^{G(x)}}{|q\omega(x)-1|^{2(n+1)}}.
\end{eqnarray}


\subsection{The case $\gamma = 1$}

For this case the adimensional variables are given by 
\begin{eqnarray}\label{omega2}
e^{-\omega}=\frac{|\mu|}{|\mu_f|},
\end{eqnarray}

\begin{eqnarray}
A^2 = \frac{q}{4\pi |\mu_f|},
\end{eqnarray}
and, $r$, $m$ and $q$ defined the same as in the expressions (\ref{variable1})--(\ref{variable3}).\\
Introducing these expressions in the (\ref{TOV2}) equation we get

\begin{eqnarray}
\frac{d\omega}{dx} -  ( q-1)\left(\frac{ q x^3 e^{-\omega}-\eta}{2q\eta - x}\right)\frac{1}{x} - \frac{2e^{\omega}\Pi}{xP^f_r} = 0. \nonumber \\ && \label{le-gamma1}
\end{eqnarray}
On the other hand, from (\ref{mp}), we obtain

\begin{eqnarray}\label{eta-gamma1}
\frac{d\eta}{dx} = x^2 e^{-\omega}.
\end{eqnarray}

Introducing the dimensionless variables in \eqref{TOV}, we can write

\begin{eqnarray}
    \frac{d\nu}{dx}=\frac{2q}{q-1}\frac{d\omega}{dx}-\frac{4 e^{\omega}\Pi}{x|\mu_f|(q-1)}.
\end{eqnarray}

Finally, following the same procedure carried out for the case $\gamma\not=1$ and using (\ref{atm})-(\ref{nrad}), we can write the Tolman mass as

\begin{eqnarray} \label{tg1}
\eta_{T} &=& \sqrt{2y-1}e^{\frac{q\omega}{q-1}+\tilde{G}}\bigg(\frac{x_{\Sigma^e}z}{2q\eta-x_{\Sigma^e}z}\bigg)^{1/2} \nonumber \\ &\times& ( qx^3_{\Sigma^e}z^3e^{-\omega}-\eta),
\end{eqnarray}
where

\begin{equation}
    \tilde{G}=\frac{2q}{P^f_r(q-1)}\int_{x}^{x_{\Sigma^e}}\frac{e^{\omega}\Pi}{\tilde{x}}d\tilde{x}.
\end{equation}\\

Equations (\ref{le1}) and (\ref{le2}) (and also (\ref{le-gamma1}) and (\ref{eta-gamma1})) form a system of two first order ordinary differential equations for the three unknown functions $\omega, \eta, \Pi$, depending on the parameters $n, q$ attached to certain boundary conditions. They correspond to the differential equations that represent the modified Lane-Emden system for the hyperbolic anisotropic polytrope and which constitutes the source of the hyperbolic vacuum solution within the horizon. Thus it is obvious that in order to proceed further with the modeling of a compact object, we need to provide additional information that depends on the specific physical problem under consideration. The fact that the  principal stresses are unequal produces an extra indeterminacy so the introduction of an additional condition to close the system is compulsory \cite{hod08,8p}. For example, in \cite{1p,2p} it was considered a particular ansatz which allowed to obtain an anisotropic model continually linked with the isotropic case \cite{Cosenza}. Another interesting choice for the local pressure anisotropy was introduced in \cite{5p, 6p} where the main idea was the additional assumption that both principal stresses satisfy polytropic equations of state. In the same way, the conformally flat \cite{3p, Herrera2001} or the vanishing complexity condition \cite{VC, 7p, bh, durgapal} have been been frequently used in the literature in order to close the system of equations.

\section{Cosenza-Herrera-Esculpi-Witten anisotropy}\label{chew}

As previously commented, in order to solve the problem of the general relativistic polytrope for anisotropic matter,  additional  information (besides (\ref{pos})) must be provided. So, in order to obtain models to illustrate our method we shall find a specific solution of the hyperbolic polytrope extending the heuristic procedure developed in \cite{Cosenza} which allows one to find anisotropic matter solutions from any known isotropic one, in the spherically symmetric case. The basic ansatz of the method is based on a specific form of the anisotropy, more specifically it is assumed that
\begin{eqnarray}
-\Pi= \Delta = P_{\perp}-P_r = C (P_r-|\mu |) \frac{\nu '}{2} r,
\label{ani1}
\end{eqnarray}
where $C$ is a parameter which measures the anisotropy. Then, using the above expression in (\ref{TOV})
\begin{eqnarray}
P'_r + (P_r-|\mu |) \frac{\nu'}{2} h = 0, 
\label{ani2}
\end{eqnarray}
with $h \equiv 1-2C$. Obviously $h = 1$ corresponds to the isotropic pressure case. We assume $h$ to be constant throughout the sphere, which of course does not imply the constancy of either pressure. Then assuming the energy density distribution of a given isotropic solution we may find the corresponding anisotropic model satisfying (\ref{ani1}). 

Now, using (\ref{nuprima}) we can write (\ref{ani1}) in a useful way,
\begin{eqnarray}\label{ani3}
\Pi=\frac{h-1}{2} \left(\frac{4\pi r^3 P_r - m}{2m-r}\right)(P_r-|\mu|). 
\end{eqnarray}

\subsection{The case $\gamma \not= 1$}
Introducing (\ref{ani3}) in (\ref{le1}) we get
\begin{eqnarray} \label{lec}
\frac{d\omega}{dx} +  h\frac{( qx^3\omega^{n+1}-\eta)(q\omega-1)}{x \left[2q(n+1)\eta -x\right]} = 0,
\end{eqnarray}
which, together with (\ref{le2}), conform the Lane-Emden system of equations.

Next, using (\ref{TMn1}) and (\ref{Gn1}), we can write the Tolman mass for this case

\begin{eqnarray} \label{TMc}
\eta_T &=& \frac{\sqrt{2y-1}}{|q\omega-1|^{(n+1)/h}}\bigg[\frac{x_{\Sigma^e}z}{2q(n+1)\eta-x_{\Sigma^e}z}\bigg]^{1/2} \nonumber \\ &\times& (qx^3_{\Sigma^e}z^3\omega^{n+1}-\eta).
\end{eqnarray}

\begin{figure*}
    \resizebox{1.0 \textwidth}{!}{
    \includegraphics{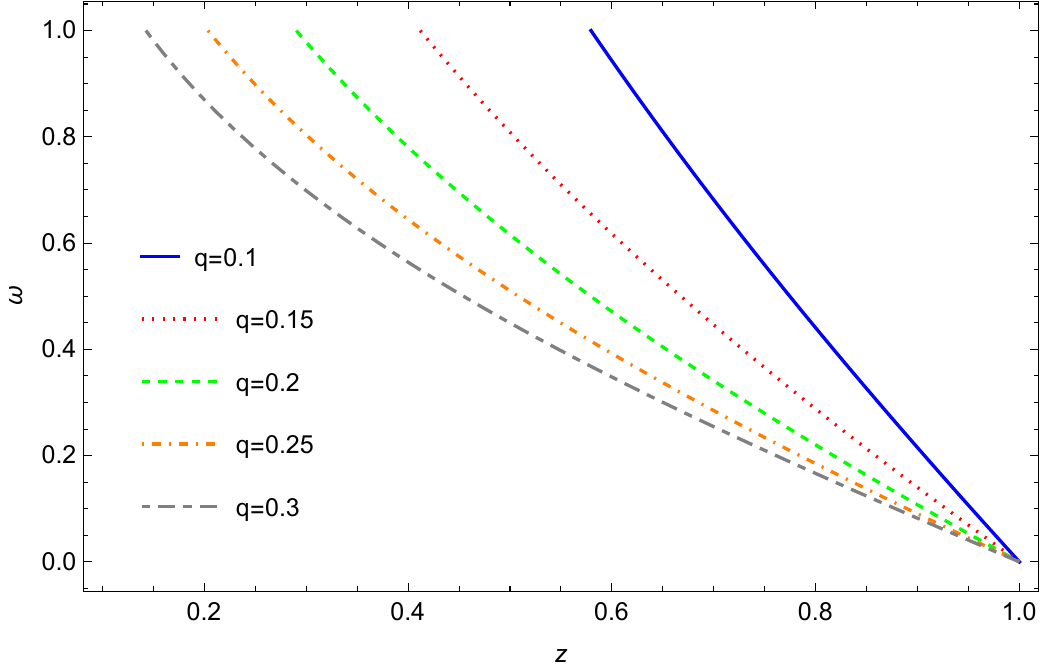} \ 
    \includegraphics{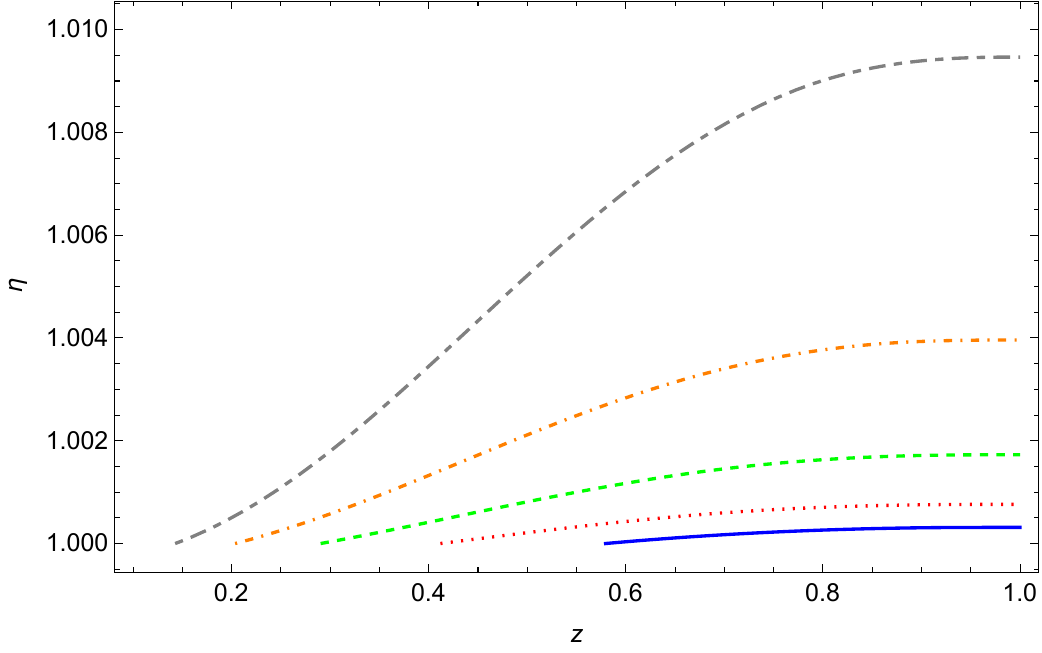}}
    \resizebox{1.0 \textwidth}{!}{
    \includegraphics{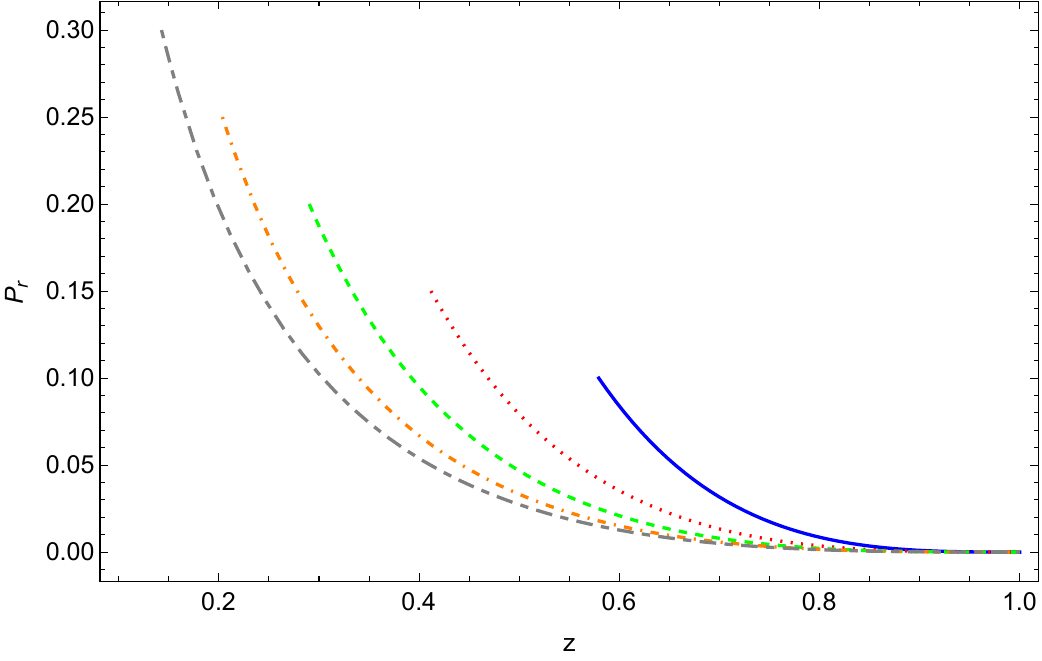} \ 
    \includegraphics{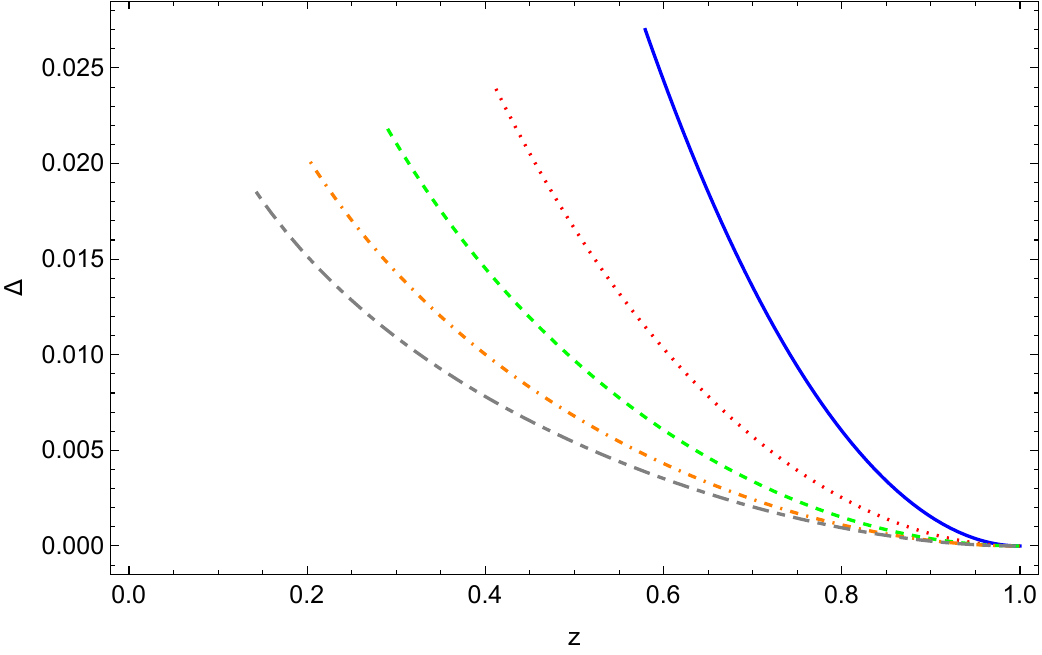}}
    \resizebox{0.5 \textwidth}{!}{
    \includegraphics{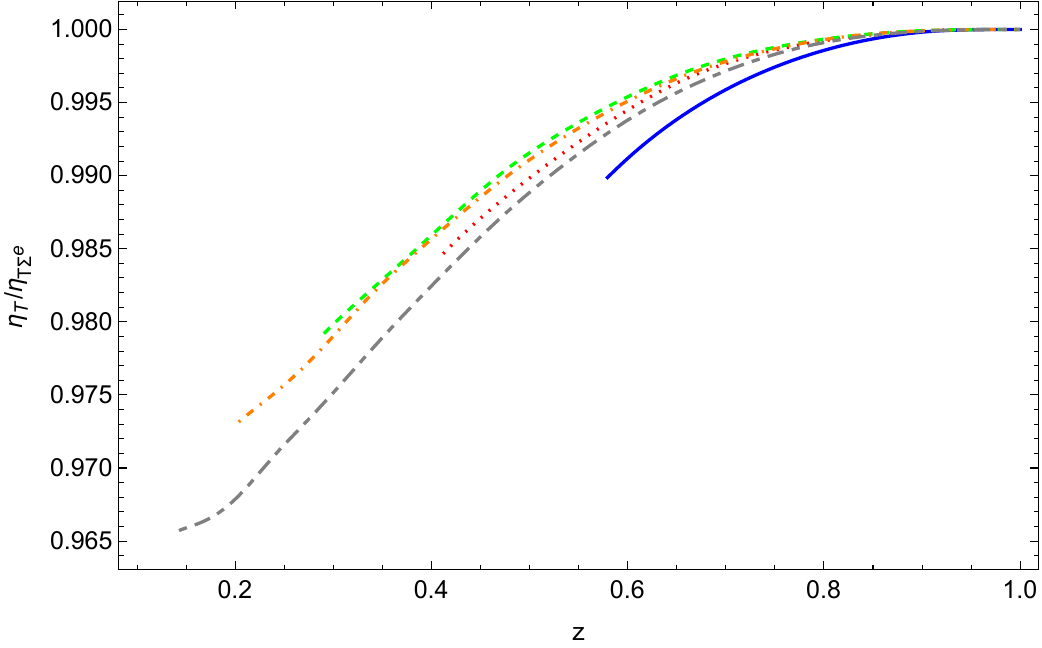} 
    } 
    \caption{General Profile for $\gamma\neq 1$ for same $n=2$ and $h=0.9$ for values $x_{\Sigma^i}=0.1$, $\omega(x_{\Sigma^i})=1$ and $\eta(x_{\Sigma^i})=1$.}
    \label{fig4}
\end{figure*}

\begin{figure*}
    \resizebox{1.0 \textwidth}{!}{
    \includegraphics{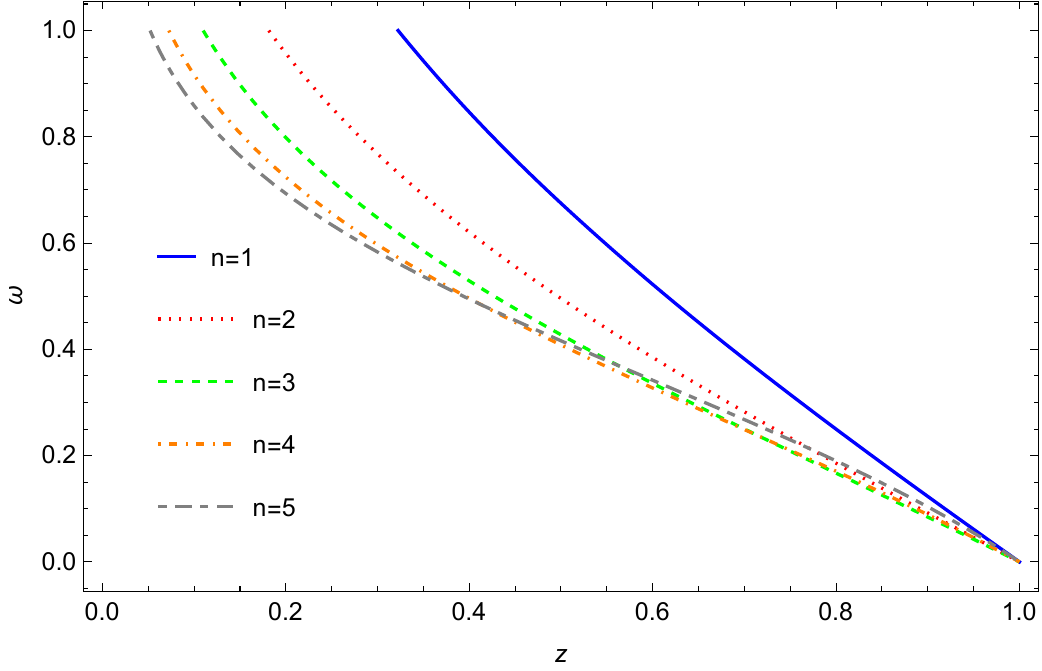} \ 
    \includegraphics{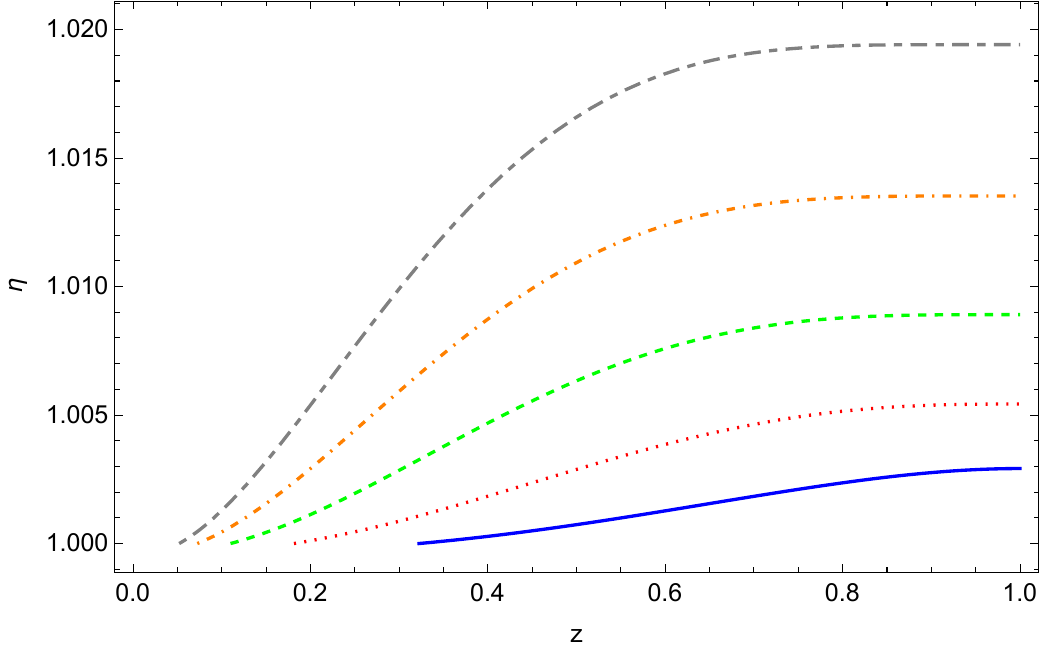}}
    \resizebox{1.0 \textwidth}{!}{
    \includegraphics{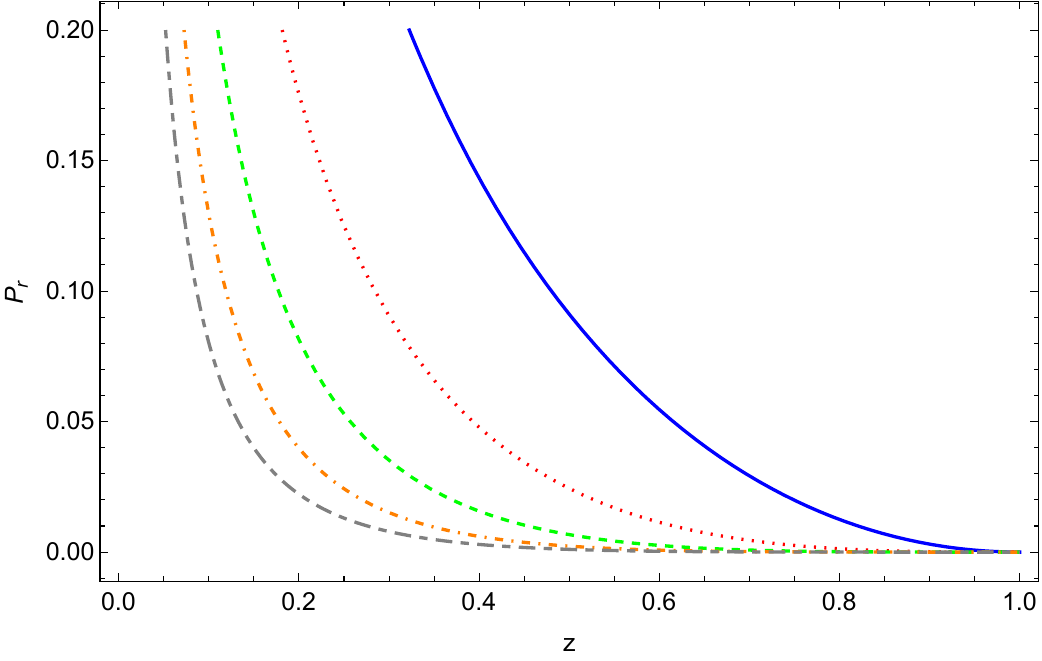} \ 
    \includegraphics{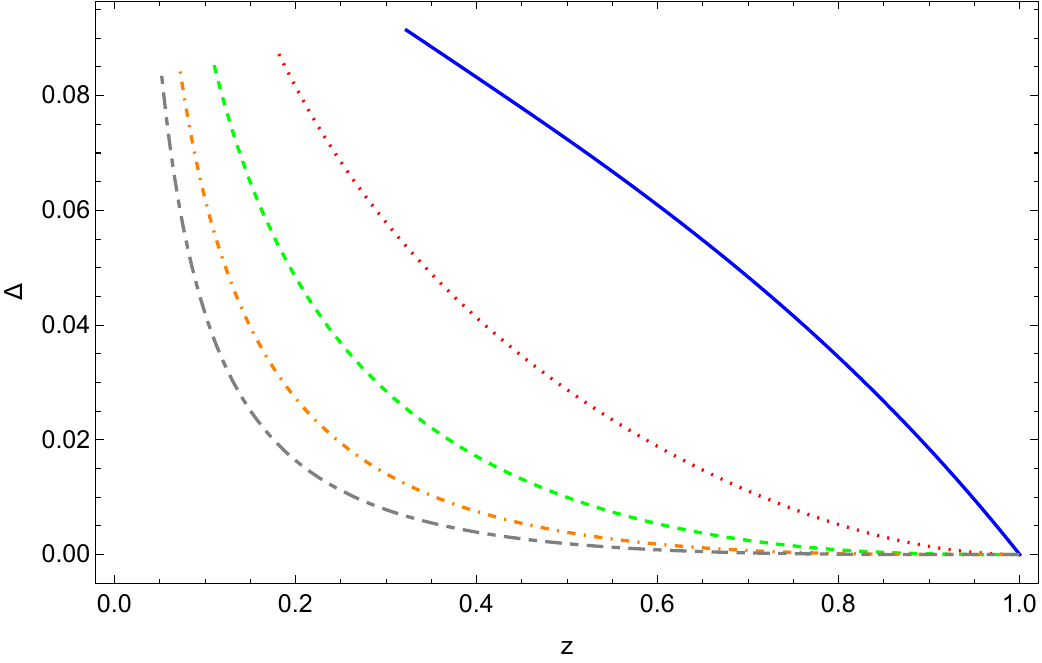}}
    \resizebox{0.5 \textwidth}{!}{
    \includegraphics{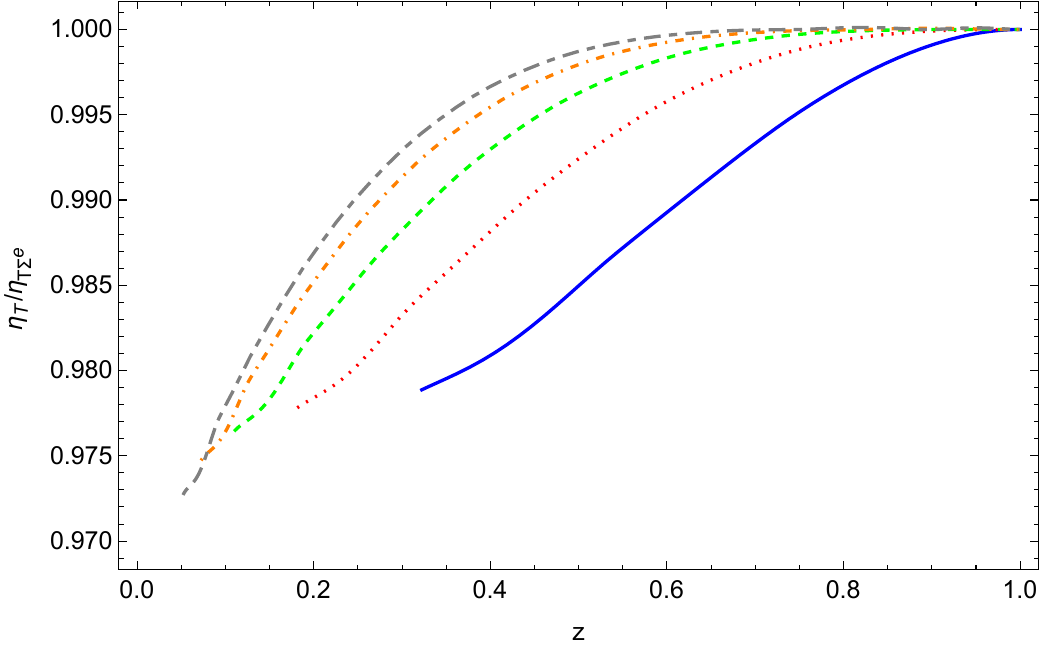} 
    } 
    \caption{General Profile for $\gamma\neq 1$ for same $h=0.6$ and $q=0.2$ for values $x_{\Sigma^i}=0.1$, $\omega(x_{\Sigma^i})=1$ and $\eta(x_{\Sigma^i})=1$.}
    \label{fig5}
\end{figure*}

\begin{figure*}
    \resizebox{1.0 \textwidth}{!}{
    \includegraphics{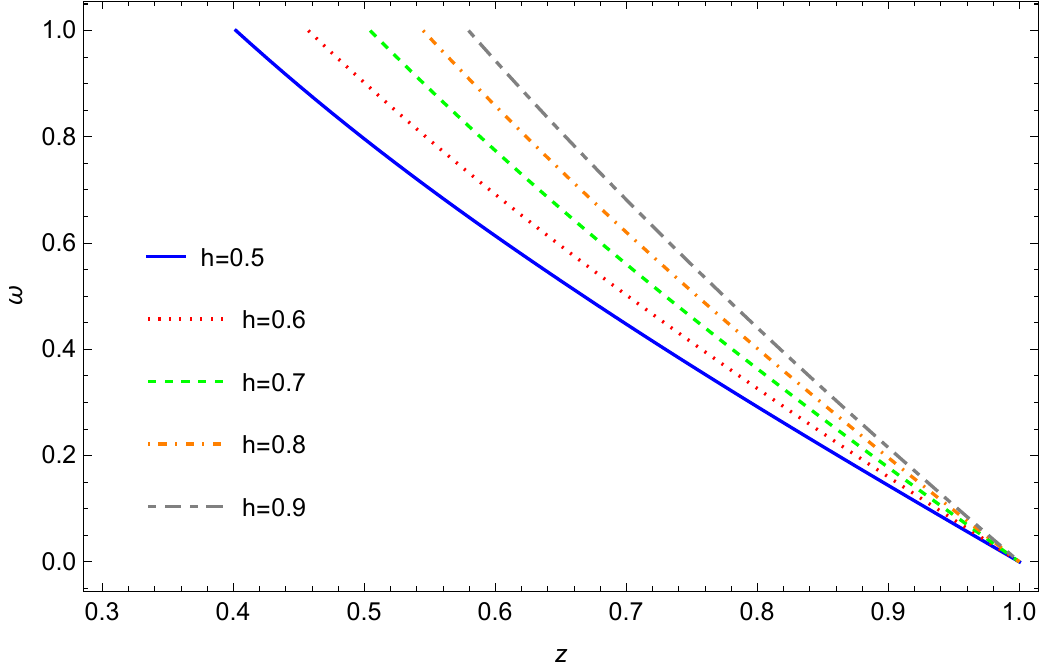} \ 
    \includegraphics{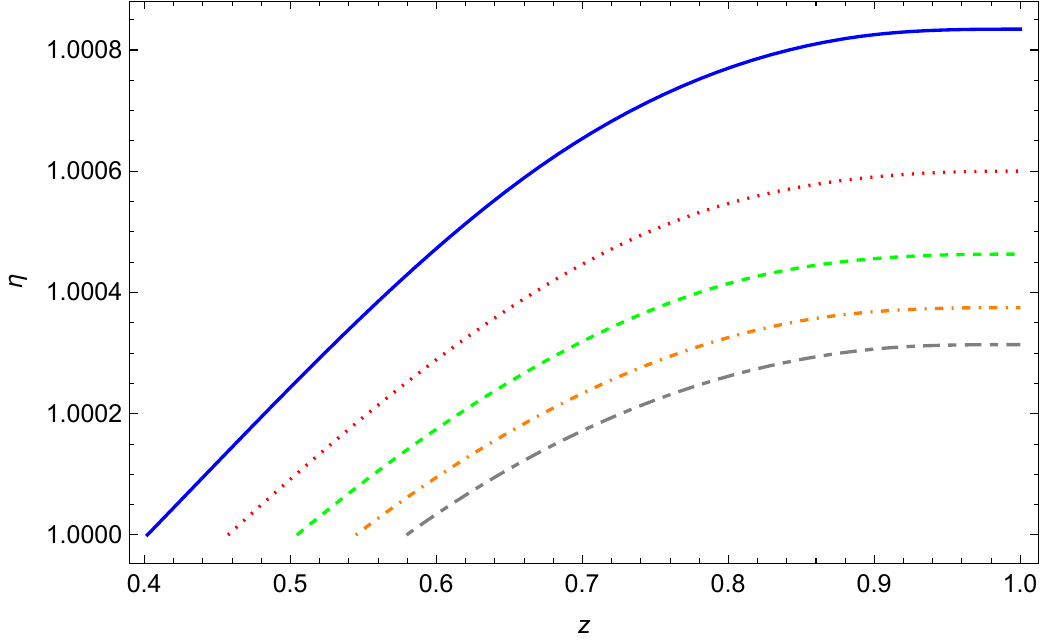}}
    \resizebox{1.0 \textwidth}{!}{
    \includegraphics{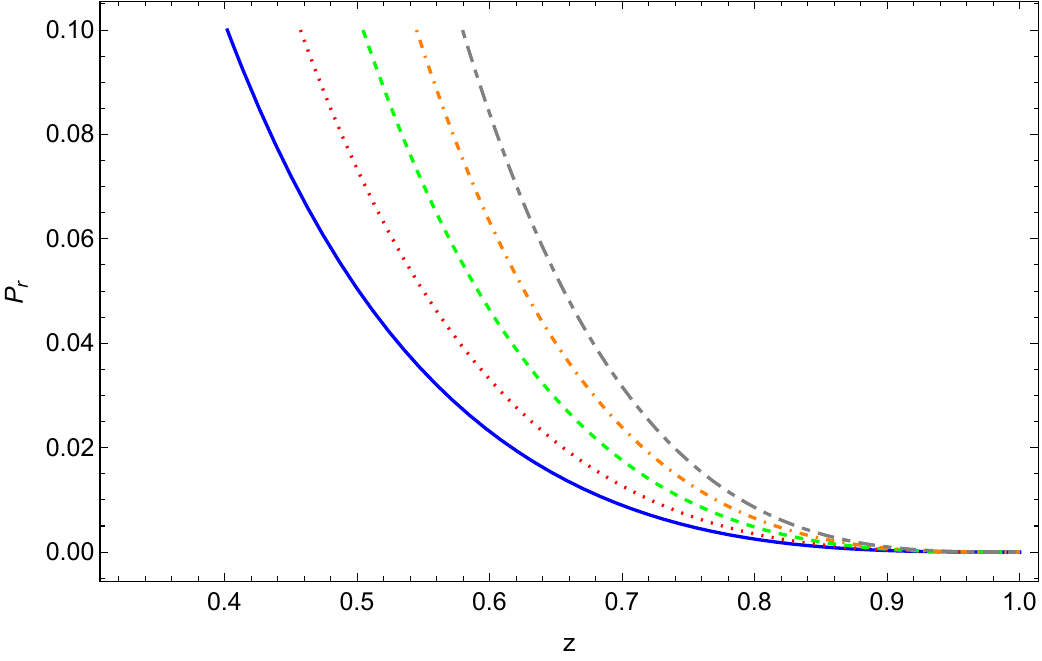} \ 
    \includegraphics{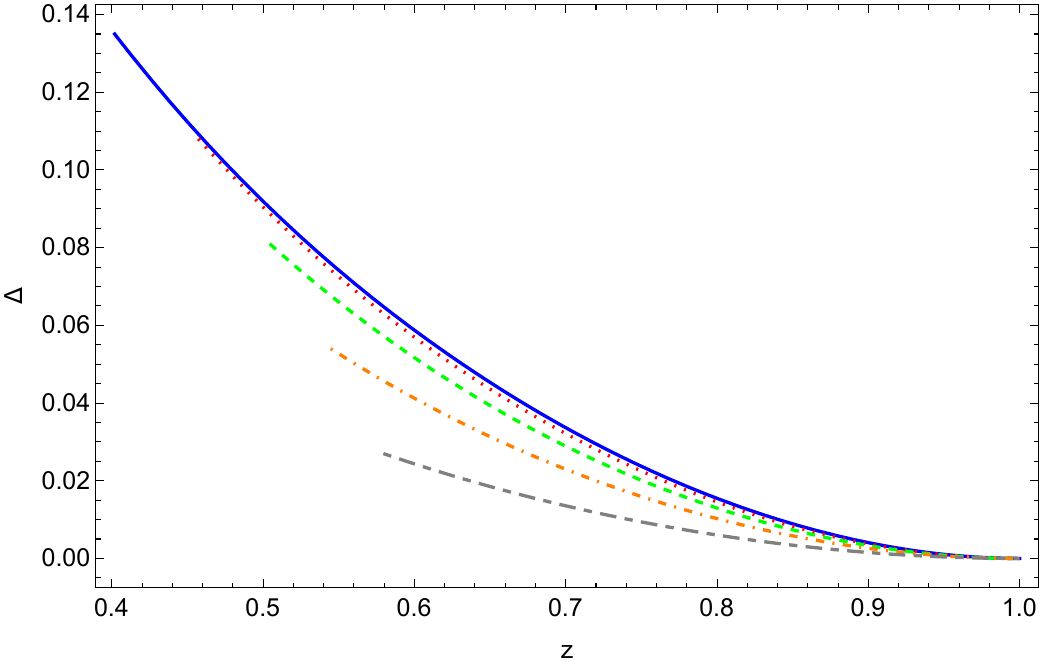}}
    \resizebox{0.5 \textwidth}{!}{
    \includegraphics{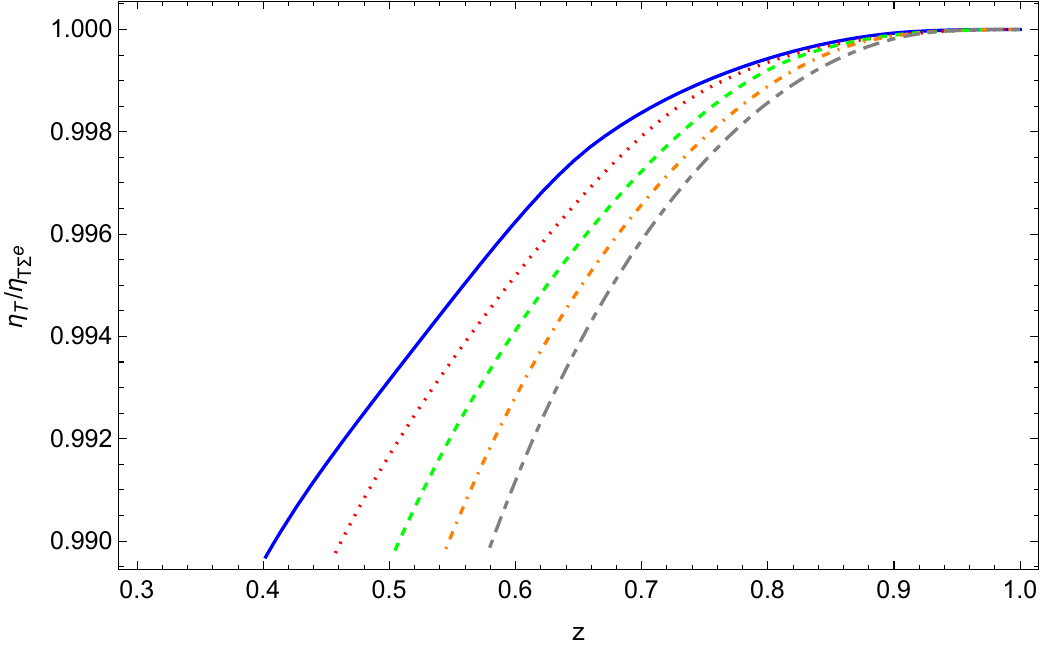} 
    } 
    \caption{General Profile for $\gamma\neq 1$ for same $n=2$ and $q=0.1$ for values $x_{\Sigma^i}=0.1$, $\omega(x_{\Sigma^i})=1$ and $\eta(x_{\Sigma^i})=1$.}
    \label{fig6}
\end{figure*}

\subsection{The case $\gamma=1$}
Introducing (\ref{ani3}) in (\ref{le-gamma1}) we obtain
\begin{eqnarray}\label{le3}
\frac{d\omega}{dx} -  h ( q-1)\left[\frac{ q x^3 e^{-\omega}-\eta}{x(2q\eta - x)}\right] = 0. 
\end{eqnarray}
which, together with (\ref{eta-gamma1}), conform the Lane-Emden system of equations.

The Tolman mass for this case can be written as

\begin{eqnarray}
\eta_{T} &=& \sqrt{2y-1}e^{\frac{q\omega}{h(q-1)}}\bigg(\frac{x_{\Sigma^e}z}{2q\eta-x_{\Sigma^e}z}\bigg)^{1/2} \nonumber \\ &\times& ( qx^3_{\Sigma^e}z^3e^{-\omega}-\eta).
\end{eqnarray}

\begin{figure*}
    \resizebox{1.0 \textwidth}{!}{
    \includegraphics{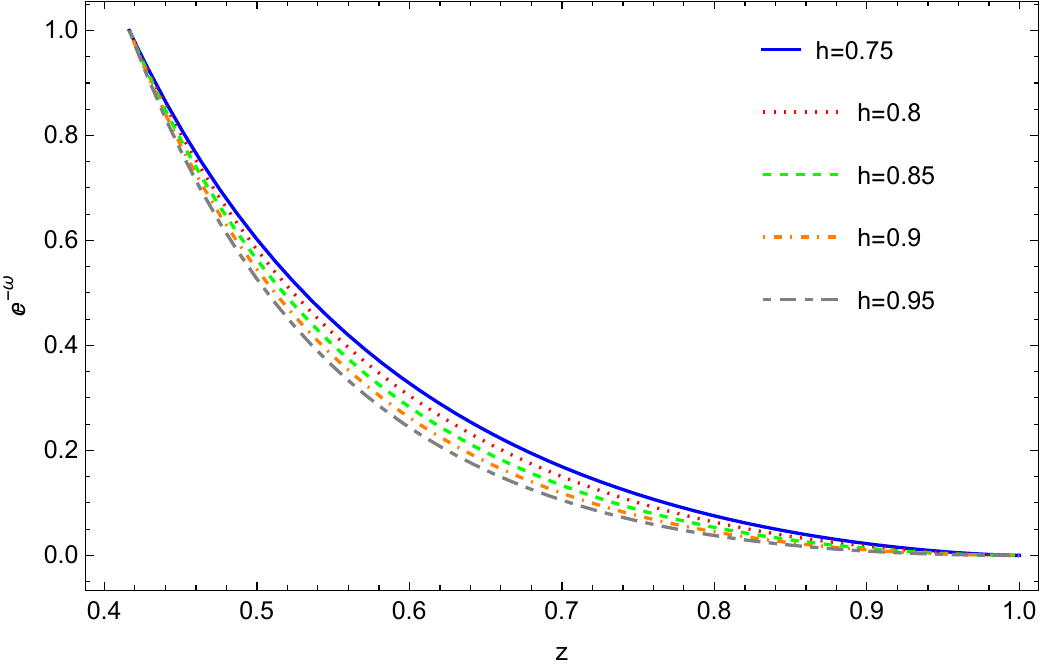} \ 
    \includegraphics{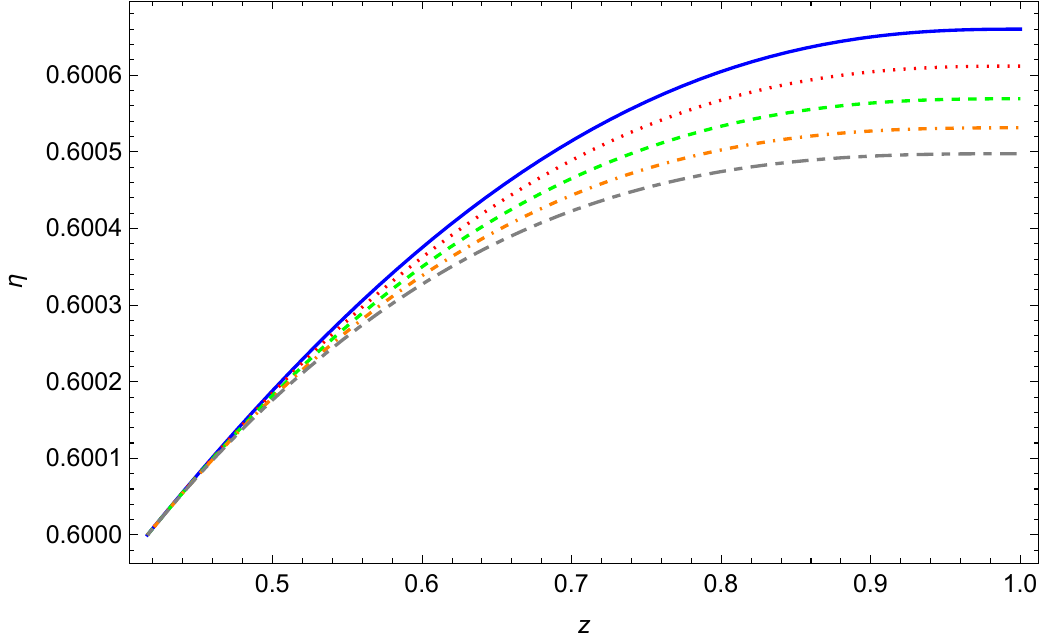}}
    \resizebox{1.0 \textwidth}{!}{
    \includegraphics{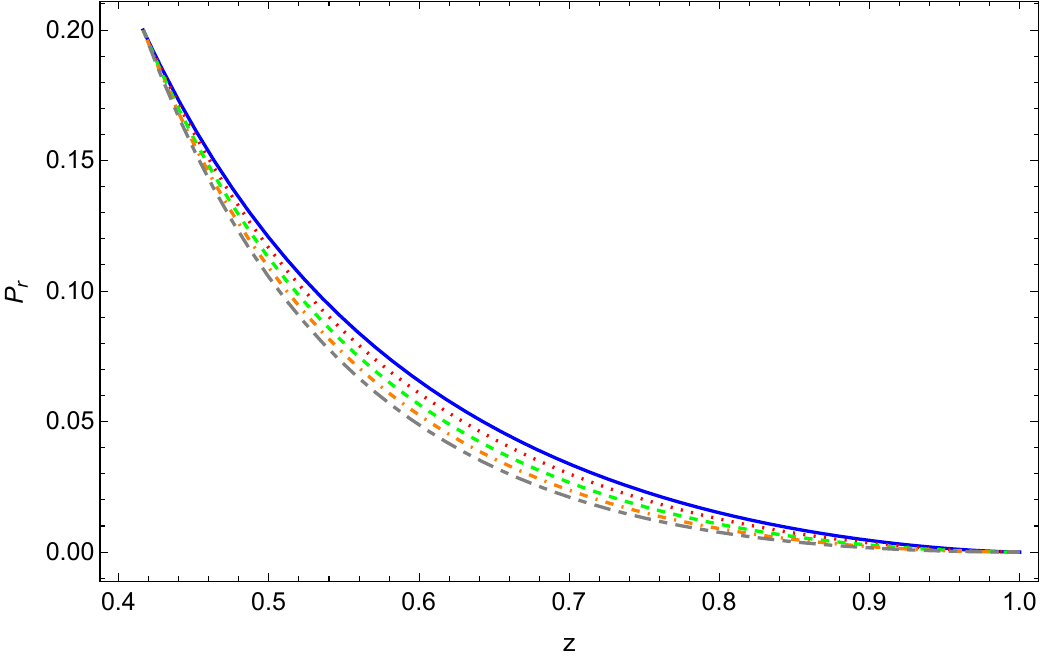} \ 
    \includegraphics{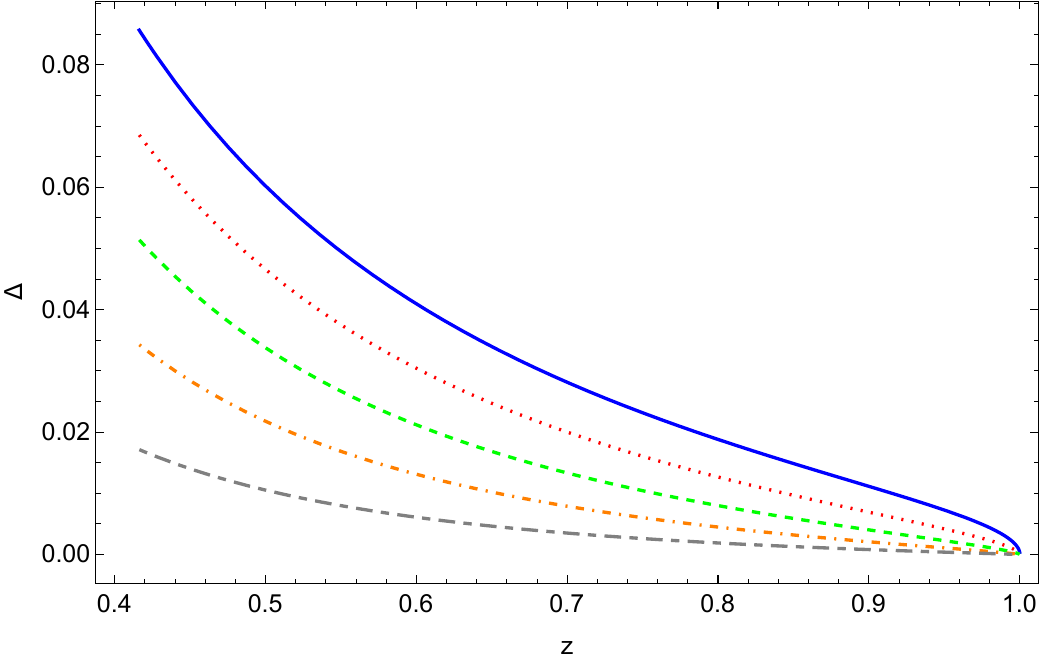}}
    \resizebox{0.5 \textwidth}{!}{
    \includegraphics{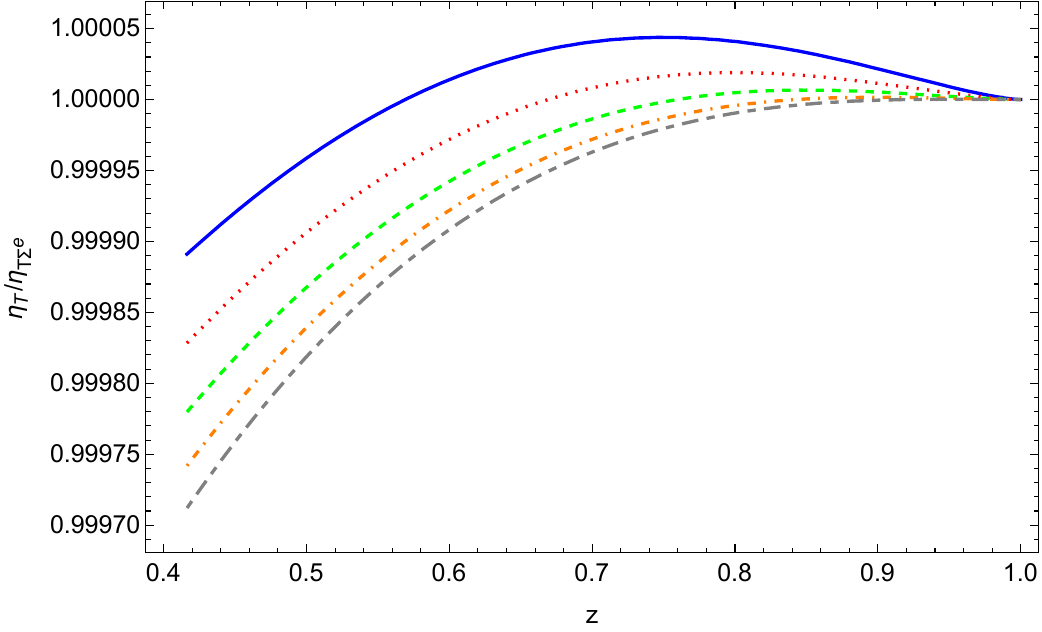} 
    } 
    \caption{General Profile for $\gamma =1$ for same $q=0.2$ for values $x_{\Sigma^i}=0.1$, $\omega(x_{\Sigma^i})=0$ and $\eta(x_{\Sigma^i})=0.6$.}
    \label{fig7}
\end{figure*}

\begin{figure*}
    \resizebox{1.0 \textwidth}{!}{
    \includegraphics{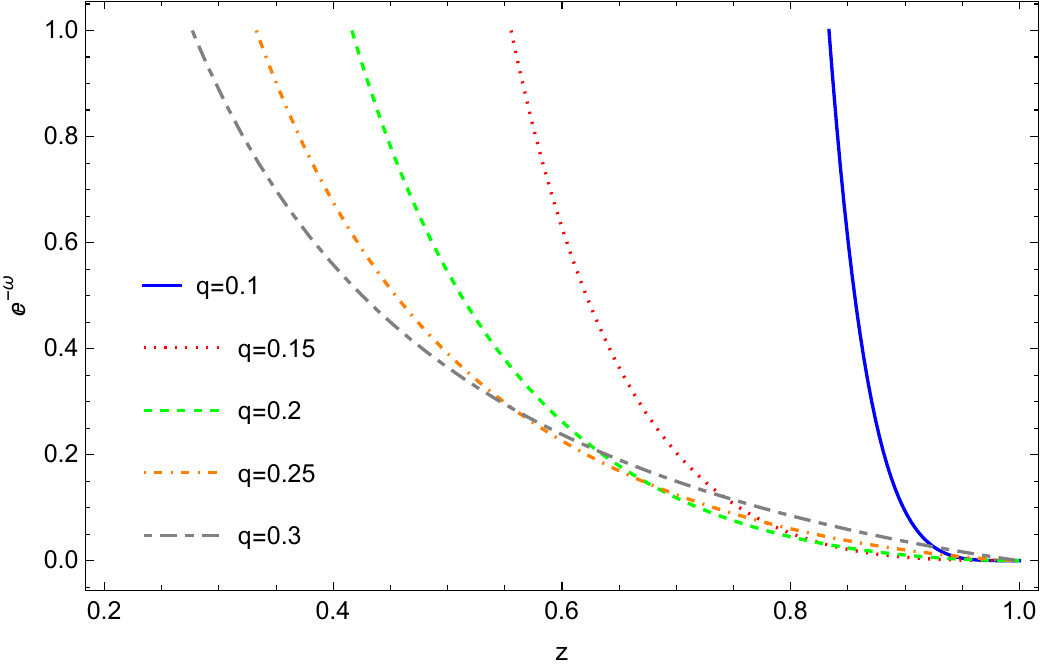} \ 
    \includegraphics{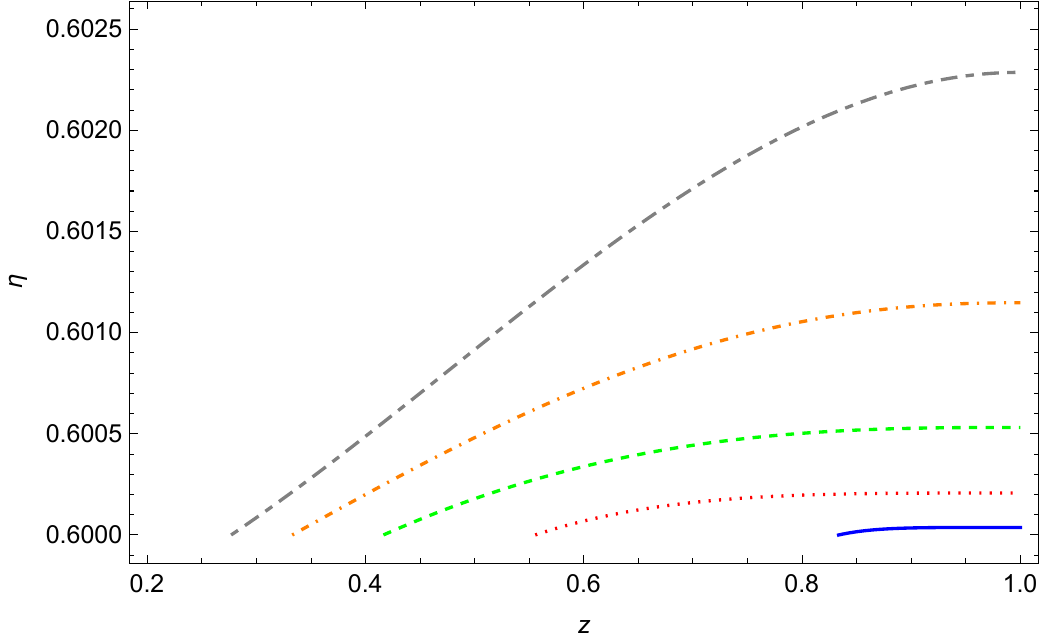}}
    \resizebox{1.0 \textwidth}{!}{
    \includegraphics{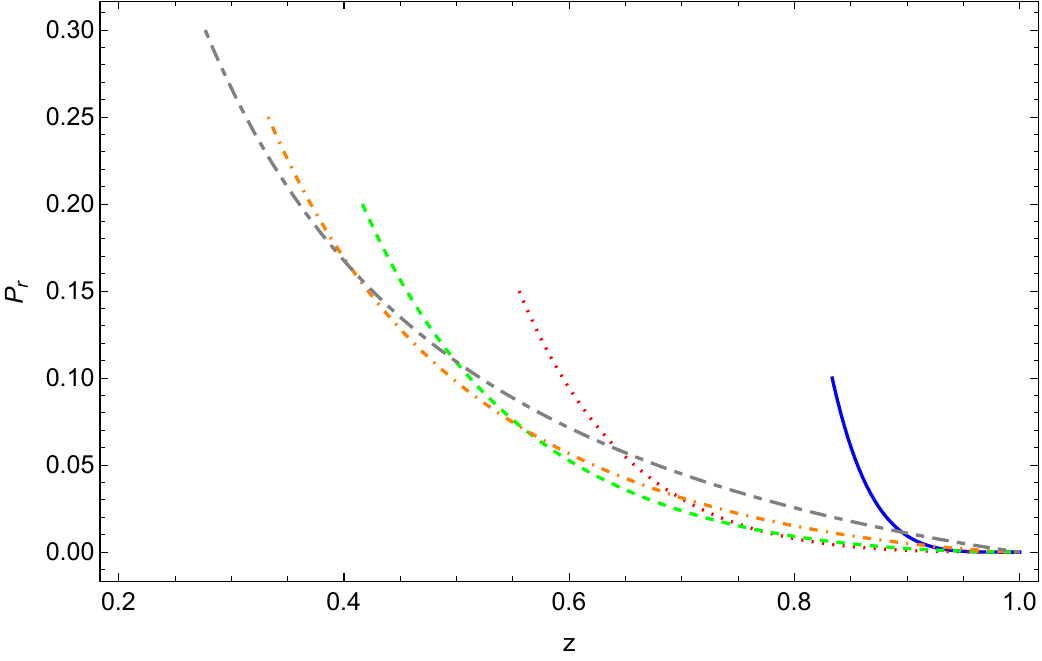} \ 
    \includegraphics{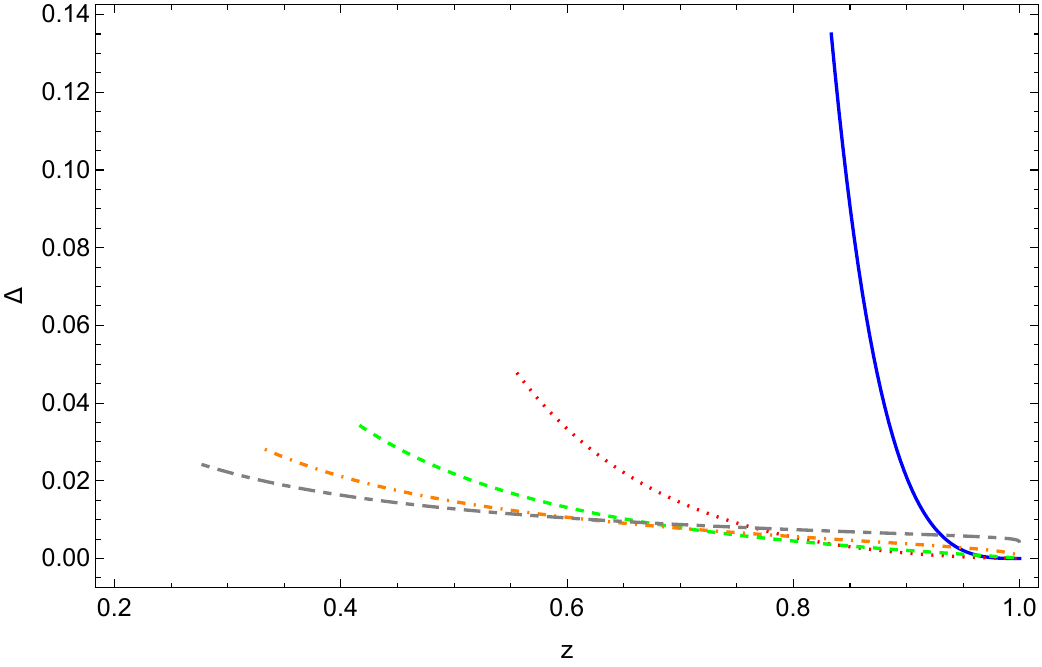}}
    \resizebox{0.5 \textwidth}{!}{
    \includegraphics{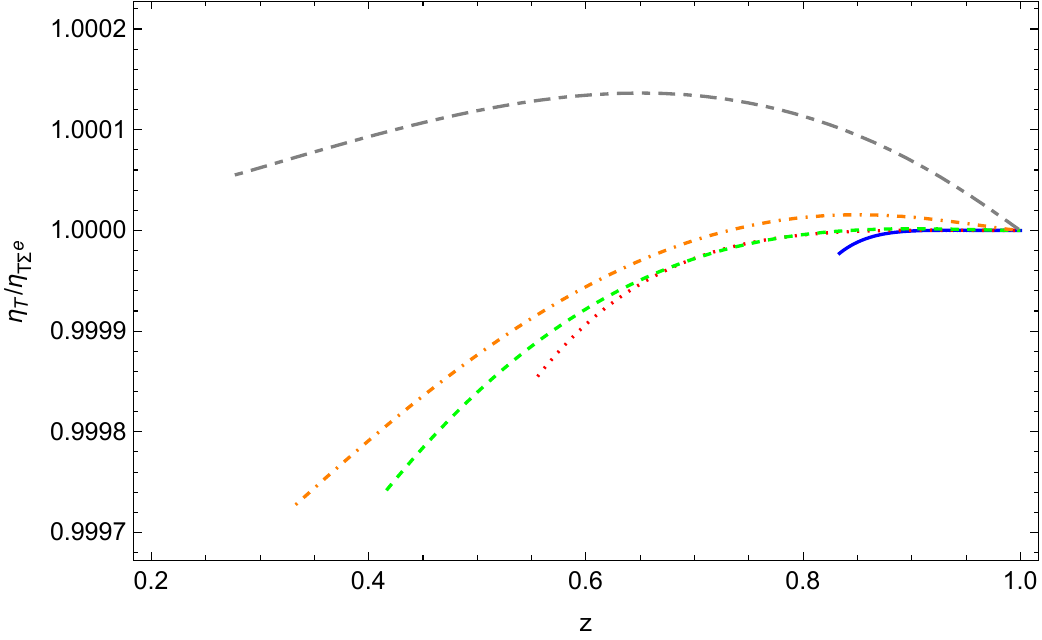} 
    } 
    \caption{General Profile for $\gamma =1$ for same $h=0.9$ for values $x_{\Sigma^i}=0.1$, $\omega(x_{\Sigma^i})=0$ and $\eta(x_{\Sigma^i})=0.6$.}
    \label{fig8}
\end{figure*}

\begin{figure}
    \centering
    \includegraphics[width=8cm]{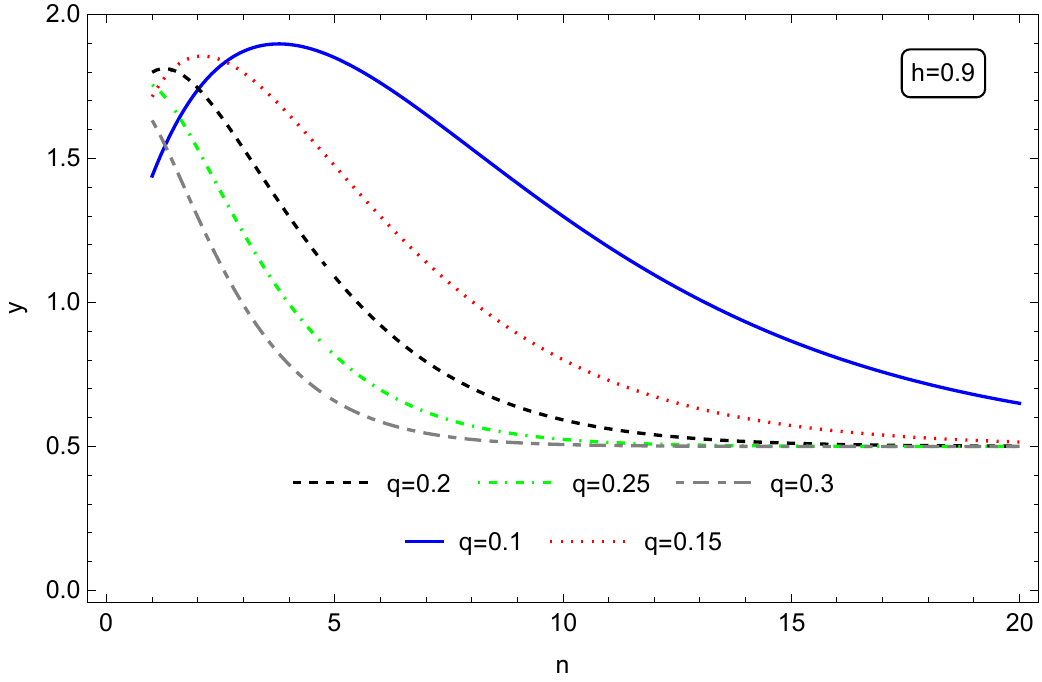}
    \caption{Degree of compactness $y$ vs. $n$ for different values of q and same $h=0.9$. Notably, when $n\to\infty$, $y\to 0.5$.}
    \label{figy}
\end{figure}


\section{Discussion}\label{discussion}

In the study of stellar configurations, the behavior of the matter sector is crucial for determining whether the solution is physically acceptable or not. So, we proceed to integrate the equations (\ref{lec}) and (\ref{le3}) numerically by exploring the set of parameters involved with the aim to study the behaviour of the matter sector, namely, the density energy, the pressures, the anisotropy, the Tolman mass and the surface potential. Specifically, in spherically symmetric spacetimes the 
density and pressures must decrease from the center to the surface, while the mass function and Tolman mass  should increase. Another important aspect is the role of anisotropy, $\Delta = P_{\perp} - P_{r}$, which, as mentioned in the introduction, naturally arises due to the processes occurring during star formation. Note that anisotropy can be either negative or positive depending on whether the radial pressure is greater or less than the tangential pressure, respectively. However, it is expected to be positive so that the force associated with it contributes to strengthening the force due to the radial pressure gradients, thus counteracting the gravitational term (otherwise, anisotropy would reinforce the gravitational force and promote a possible collapse of the configuration). On the other hand, whether
$\Delta$ is increasing or decreasing will depend on the behavior of the radial pressure gradient, as discussed further below. The previously described behavior should also hold in hyperbolic symmetry with some subtle differences. For example, since the energy density is negative, it is expected that its norm should be monotonically decreasing towards the surface. On the other hand, the Tolman mass can be negative in hyperbolic symmetry, but the normalized Tolman mass is always positive. However, whether the normalized Tolman mass is monotonically increasing towards the surface depends on the sign of the integrand $-|\mu| + P_r + 2P_\perp$ in equation (\ref{mT1}), since if there is a change in the sign of the integrand, the normalized Tolman mass would reach a local maximum at some point. Of course, a sign change would imply that in that region $\mu + P_r + 2P_\perp > 0$, which would mean that the solution would satisfy the strong energy condition there.

As discussed in the previous section, in this work, the information associated with the energy density is encoded in the quantity $\omega$ given by definitions (\ref{omega1}) (for $\gamma \ne 1$) and (\ref{omega2}) (for $\gamma = 1$). The radial pressure is given by the polytropic equation of state (\ref{Pr}), and the information about the mass function and Tolman mass is contained in the quantities \(\eta\) and \(\eta_{T}\), respectively, given by expressions (\ref{variable2}) and (\ref{TMn1}) for the case  $\gamma \ne 1$ and the respective expressions (\ref{tg1}) for the case $\gamma = 1$. To find these quantities, we must numerically solve the differential equations given by (\ref{le1}) and (\ref{le-gamma1}) for $\gamma \ne 1$ and $\gamma = 1$, respectively, in combination with the equation for the mass function given by (\ref{le2}) and (\ref{eta-gamma1}) for each case. This numerical integration is performed by fixing points in the parameter space $(q, n, h)$. Figures \ref{fig4}-\ref{fig6} show the plots of $\omega$, $\eta$, $P_{r}$, $\Delta$, and $\eta_{T}/\eta_{T\Sigma^{e}}$ for $\gamma \ne 1$. Figure \ref{fig4} presents the results of numerical integration for different values of $q$ with $(h, n)$ fixed. Figure \ref{fig5} shows the results for different values of $n$ with $(h, q)$ fixed, and Figure \ref{fig6} shows the results for varying $h$ . All these results share a common feature: the norm of the energy density and radial pressure decrease towards the exterior surface, the mass function and the (normalized) Tolman mass increase towards the surface, and the anisotropy is positive, as expected. However, a couple of comments are in order. First, note that although the Tolman mass is negative, the plot we show here is positive given we are dealing with the normalized quantity $\eta_{T}/\eta_{T\Sigma^{e}}$.
Second, there is a very peculiar behavior of the anisotropy that we want to highlight: in all cases, it monotonically decreases towards the surface. This can be easily understood from the behavior of the TOV equation (\ref{TOV2}). The first term corresponds to pressure gradients due to nuclear reactions inside the compact object that counteract the (possible) gravitational collapse. Since the pressure is decreasing, this quantity is negative. The second term, proportional to $P_{r} - |\mu|$, corresponds to the gravitational force and is positive, while the last term is proportional to $-\Delta$. Note that, due to the polytropic equation of state, the second term decreases as we approach the outer surface until it vanishes at $r_{\Sigma^{e}}$ (the radial pressure must vanish at the outer surface, and since pressure is proportional to a potential of the energy density, it must also vanish for consistency). In this sense, only the first and third terms remain. Now, if we observe the behavior of the radial pressure, we notice that it has a minimum at the surface, so its derivative vanishes there. Consequently, anisotropy must also vanish at the surface in consistency with the hydrostatic equilibrium given by (\ref{TOV2}). The results of the numerical integration for the case $\gamma = 1$
are shown in Figures \ref{fig7} and \ref{fig8}, varying $q$ and $h$, respectively. In both plots, the behavior is similar to that discussed in the previous paragraph for $\gamma \ne 1$ except for the normalized Tolman mass function which has a local maximum near the exterior surface indicating the fulfilling of the strong energy condition there. Although due to this fact one might mistakenly think that instabilities are generated by the accumulation or the increasing of the ``active" gravitational mass towards interior regions, but the fact actually occurs in middle layers (more towards the outside) of the distribution between the interior and exterior surfaces, rather avoiding collapse and stabilizing the object.

Figure \ref{figy} shows the compactness of the star, which corresponds to the ratio of the total mass to the outer radius, i.e., $y = M / r_{\Sigma^{e}}$. The graph shows how compactness changes as a function of the polytropic index $n$ with $h$ fixed and for different values of $q$. It is notable that
$y > 0.5$ as opposed to what occurs in the spherical case. This is expected, as the hyperbolic vacuum corresponds to $\frac{2M}{r_{\Sigma^{e}}} - 1 > 0$. It is also noteworthy that as $n \to \infty$, the compactness asymptotically tends to its limit value of 0.5. In general, $y$ reaches a maximum for a certain value of $n$ and then decreases. As $q$ decreases, the value of $n$ at which the maximum $y$ is reached is larger, and the maximum shifts to the right with decreasing parameter $q$, which is an interesting behavior indicating that there is a correlation between $q$ and compactness. Note that, although for the hyperbolic case, compactness has a lower bound given by the metric parametrization, it also has an upper bound, which in this case is $y \sim 2$. This point is important to note since, in the spherical case, the finiteness of $y$ is ensured by the upper bound, but nothing guarantees that $y$ must be finite for all values of $n$ in the hyperbolic case. Initially, one might think that since $y > 0.5$, compactness could become arbitrarily large, but this is not the case. In fact, this would lead to a conceptual problem since we would not know how to interpret what a star with infinitely large compactness is, other than a black hole.

Finally, it is worth noticing that for the hyperbolic polytrope the parameter $q$ does not have the simple meaning that it had for the usual spherically symmetrical polytrope theory considered the stiffness at the centre of the matter distribution, indicating how relativistic is the polytrope (see for example the references \cite{2p,6p,LN3,efl}). In fact, $q_c = \frac{P_{rc}}{\rho_c}$ allowed us to obtain the Newtonian limit and $q_c =1$ is the extreme (maximal) value that corresponds to the stiff equation of state $P_r=\rho$, which is believed to describe ultradense matter \cite{zh}. In the hyperbolic case this parameter does not have the same interpretation, must be defined by the quotient of the respective thermodynamic quantities between the two boundary surfaces of the fluid by means of (\ref{variable3}).


\section{conclusions}

Our purpose in this work was to carry out a complete study for an internal solution that obeys the equation of state of a polytrope, for hyperbolic symmetry, located in the region inner to the horizon. The fact of using a polytropic equation of state to model compact objects has been carried out for several years producing very interesting candidates. Also, due to the great relevance that anisotropic internal solutions have acquired in recent years in the structure of self–gravitating objects, and by the fact that polytropes represent fluid systems with a wide range of applications in astrophysics (e.g. Fermi fluids, super-Chandrasekhar white dwarfs), we have described hereby a general framework for modeling general relativistic polytropes using hyperbolic symmetry. We have studied an interior solution for a hyperbolic polytrope, which, to the best of our knowledge, is the first time such an analysis has been conducted. The fluid is defined between an inner radius, which is coupled to an isotropic fluid with constant density, and an outer radius, which is coupled to a hyperbolic vacuum that, in turn, connects with the Schwarzschild vacuum. In order to close the system and integrate the obtained set of equations, we need to provide further information about the anisotropy inherent to the problem under consideration. For doing
so, we employed the Cosenza-Herrera-Esculpi-Witten anisotropy ansatz and performed numerical integration, allowing us to specific modeling. For these models we also calculated the Tolman mass, whose behavior allows one to understand some of their peculiar features.

To achieve the objective of this work the requirement of staticity is crucial and in that context we have used the scheme presented by L. Herrera with hypertribolic symmetry \cite{Hyperbolic}. From a physical point of view, the existence of a static solution would be expected over the whole space-time, as an equilibrium final state of a physical process is expected to be static. With an analytical extension of the Schwarzschild solution this would not be possible. The Schwarzschild horizon is also a Killing horizon, implying that the time-like Killing vector outside the horizon becomes space-like inside it. If we recall that a static observer is one whose four-velocity is proportional to the Killing time-like vector, it follows that no static observers can be defined inside the horizon. In this work, we have considered the whole spacetime. However, instead of sacrifcing the staticity in the region inside the horizon, we keep the time independence but change the spatial symmetry; crucial fact that allows us the necessarily static description of a polytropic solution within the horizon.

The results demonstrate that energy density (in absolute value) and radial pressure decrease from the interior surface $\Sigma^i$ to the exterior one $\Sigma^e$, while the mass function and normalized Tolman mass increase. Anisotropy remains positive throughout, but its magnitude diminishes as one approaches the exterior surface, consistent with the hydrostatic equilibrium. The study also explores the star's compactness, defined as the ratio of total mass to the outer radius $r_{\Sigma^e}$. In the hyperbolic case, compactness is observed to be greater than 0.5, which contrasts with the spherical case. Compactness reaches a maximum value for a specific range of the polytropic index before decreasing. Despite the higher compactness observed in the hyperbolic case, it remains bounded, thus avoiding unrealistic scenarios such as infinite compactness, which would lead to conceptual issues similar to those encountered with black holes.

We have discussed the possibility of interpreting our solution as a type of gravastar in the sense of Mazzur and Mottola \cite{Mazur:2001fv}, with the difference that instead of the Schwarzschild interior solution tending toward the de Sitter solution, it coexists with the hyperbolic polytrope. It is important to highlight that we used the Cosenza-Herrera-Esculpi-Witten anisotropy for convenience in this work. Clearly, other types of anisotropy are possible and are left for future studies. However, it should be noted that we attempted a solution where anisotropy is derived from the conformally flat condition, but the results were not entirely satisfactory due to the behavior of the quantities involved. Perhaps a more precise numerical analysis or a more thorough exploration of the parameter space is needed. 

As we have mentioned, other polytropic solutions can be contemplated where the system of field equations closes using different considerations for the anisotropy: the vanishing complexity \cite{VC,VC2}, the double polytrope solution studied in \cite{6p} or even explore the possibility of having a Class I-Karmakar hyperbolic polytropic solution \cite{karmarkar,7p}. On the other hand it is well known (see for example \cite{LN3}), that implementing the polytropic equation of state leads to a singular tangential sound velocity at the surface of the distribution for $n>1$ in the case of spherical symmetry. One way to address this problem is to consider a modification of the polytropic equation of state, known as the ``master" polytropic equation of state \cite{LN3,LN2,efl}, which is given by $P_r = K_r \rho^{1+1/n_r}+ \tilde{\alpha}_r \rho - \tilde{\beta}_r$, where $\tilde{\alpha}_r$ and $\tilde{\beta}_r$ are constants, allowing to model material configurations where the density is discontinuous at the boundary. It is worth mentioning that this scheme has been used to describe several cosmological scenarios as the construction of early universe, the study of compact anisotropic charged objects \cite{45} and to find mathematical models of compact objects incorporating the radiation factor \cite{46}. Although the purpose of this work is not to model any particular astrophysical object, these extensions (that generate wide range parameter space) raise the possibility of connecting the emerging results with real astrophysical compact objects.

\section{Acknowledgements}

 P.L is grateful with ANID/ POSTDOCTORADO BECAS CHILE/ 2022 - 74220031. P. L was supported in part by a grant from the Gluskin She /Onex Freeman Dyson Chair in Theoretical Physics and by Perimeter Institute. Research at Perimeter Institute is supported in part by the Government of Canada through the Department of Innovation, Science and Industry Canada and by the Province of Ontario through the Ministry of Colleges and Universities. E. F. is grateful for / MINCYT-CDCH-UCV/ 2024 - and acknowledges support from Consejo de Desarrollo Cientfico y Humanstico - Universidad Central de Venezuela in part by a grant entitled Study of compact stellar configurations composed of spherically symmetric and hyperbolic, static and anisotropic relativistic fluids in the context of General Relativity. E. C. is funded by the Beatriz Galindo contract BG23/00163 (Spain). E. C. acknowledge Generalitat Valenciana through PROMETEO PROJECT CIPROM/2022/13.


\begin{thebibliography}{0}%
\makeatletter
\providecommand \@ifxundefined [1]{%
 \@ifx{#1\undefined}
}%
\providecommand \@ifnum [1]{%
 \ifnum #1\expandafter \@firstoftwo
 \else \expandafter \@secondoftwo
 \fi
}%
\providecommand \@ifx [1]{%
 \ifx #1\expandafter \@firstoftwo
 \else \expandafter \@secondoftwo
 \fi
}%
\providecommand \natexlab [1]{#1}%
\providecommand \enquote  [1]{``#1''}%
\providecommand \bibnamefont  [1]{#1}%
\providecommand \bibfnamefont [1]{#1}%
\providecommand \citenamefont [1]{#1}%
\providecommand \href@noop [0]{\@secondoftwo}%
\providecommand \href [0]{\begingroup \@sanitize@url \@href}%
\providecommand \@href[1]{\@@startlink{#1}\@@href}%
\providecommand \@@href[1]{\endgroup#1\@@endlink}%
\providecommand \@sanitize@url [0]{\catcode `\\12\catcode `\$12\catcode `\&12\catcode `\#12\catcode `\^12\catcode `\_12\catcode `\%12\relax}%
\providecommand \@@startlink[1]{}%
\providecommand \@@endlink[0]{}%
\providecommand \url  [0]{\begingroup\@sanitize@url \@url }%
\providecommand \@url [1]{\endgroup\@href {#1}{\urlprefix }}%
\providecommand \urlprefix  [0]{URL }%
\providecommand \Eprint [0]{\href }%
\providecommand \doibase [0]{http://dx.doi.org/}%
\providecommand \selectlanguage [0]{\@gobble}%
\providecommand \bibinfo  [0]{\@secondoftwo}%
\providecommand \bibfield  [0]{\@secondoftwo}%
\providecommand \translation [1]{[#1]}%
\providecommand \BibitemOpen [0]{}%
\providecommand \bibitemStop [0]{}%
\providecommand \bibitemNoStop [0]{.\EOS\space}%
\providecommand \EOS [0]{\spacefactor3000\relax}%
\providecommand \BibitemShut  [1]{\csname bibitem#1\endcsname}%
\let\auto@bib@innerbib\@empty
\end{thebibliography}%


\begin{thebibliography}{99}

\bibitem{Herrera:2018mzq}

L.~Herrera and L.~Witten,
Adv. High Energy Phys. \textbf{2018} (2018), 3839103
doi:10.1155/2018/3839103
[arXiv:1806.07143 [gr-qc]].


\bibitem{Rosen:1970nvr}
N.~Rosen,
doi:10.1007/978-1-4684-0721-1\_13

\bibitem{Herrera:2020bfy}
L.~Herrera, A.~Di Prisco, J.~Ospino and L.~Witten,
Phys. Rev. D \textbf{101} (2020) no.6, 064071
doi:10.1103/PhysRevD.101.064071
[arXiv:2002.07586 [gr-qc]].

\bibitem{Hyperbolic} L. Herrera, A. Di Prisco and J. Ospino, {\it Phys. Rev. D} {\bf 103}, 024037 (2021).

\bibitem{cha} S. Chandrasekhar, {\it An Introduction to the Study of Stellar Structure} (University of Chicago, Chicago, 1939).

\bibitem{3} R. Kippenhahn and A. Weigert, {\it Stellar Structure and Evolution} (Springer Verlag, Berlin, 1990).

\bibitem{7b}  C. Hansen and  S. Kawaler {\it Stellar Interiors: Physical Principles, Structure and
Evolution} (Springer Verlag, Berlin, 1994).

\bibitem{14} L. Herrera and N. O. Santos, {\it Phys. Rep.} {\bf 286},  53 (1997).

\bibitem{1p} L. Herrera and W. Barreto, {\it Phys. Rev. D} {\bf 87}, 087303, (2013).

\bibitem{5p} G. Abell\'an, E. Fuenmayor and L. Herrera, {\it Phys. Dark Univ.} {\bf 28}, 100549  (2020).

\bibitem{8p} G. Abell\'an, P. Bargue\~no, E Contreras and E. Fuenmayor, {\it Int. J. Mod. Phys. D} {\bf 29} 2050082 (2020).

\bibitem{LHP} L. Herrera, Phys. Rev. D {\bf 101}, 104024 (2020).

\bibitem{4a} R. Tooper, {\it Astrophys. J.} {\bf 140}, 434 (1964).

\bibitem{4b} R. Tooper, {\it Astrophys. J.} {\bf 142}, 1541 (1965).

\bibitem{4c} R. Tooper, {\it Astrophys. J.} {\bf 143}, 465 (1966).

\bibitem{5} S. Bludman, {\it Astrophys. J.} {\bf 183}, 637 (1973).

\bibitem{6} U. Nilsson and C. Uggla, {\it Ann. Phys.} {\bf 286}, 292 (2000).

\bibitem{7} H. Maeda, T. Harada,  H. Iguchi and N. Okuyama, {\it Phys. Rev. D} {\bf 66}, 027501 (2002).

\bibitem{11} L. Herrera and W. Barreto, {\it Gen. Relativ.  Gravit.} {\bf 36}, 127 (2004).

\bibitem{12} X. Y. Lai and R. X. Xu, {\it Astropart. Phys.} {\bf 31},  128 (2009).

\bibitem{13} S. Thirukkanesh and F. C. Ragel, {\it Pramana J. Phys.} {\bf 78},  687 (2012).

\bibitem{2p} L. Herrera and W. Barreto, {\it Phys. Rev. D} {\bf 88}, 084022, (2013).

\bibitem{3p} L. Herrera, A. Di Prisco,  W. Barreto and  J. Ospino, {\it Gen. Relativ. Gravit.} {\bf 46},  1827 (2014).

\bibitem{N1} S.A. Ngubelanga, S.D. Maharaj, {\it Eur. Phys. J. Plus} {\bf 130}, 211 (2015).

\bibitem{H1} T. Harko, M.K. Mak, {\it Astrophys. Space Sci.} {\bf 361}, 283 (2016).

\bibitem{6p} G. Abell\'an, E. Contreras, E. Fuenmayor and L. Herrerra, {\it Phys. Dark Univ.} {\bf 30}, 100632  (2020).

\bibitem{B1} M.Z. Bhatti, Z. Tariq, {\it Phys. Dark Univ.} {\bf 28}, 100482 (2020).

\bibitem{M1} S.A. Mardan, A. A. Siddiqui, I. Noureen and R.N. Jamil, {\it Eur. Phys. J. Plus} {\bf 135}, 3 (2020).

\bibitem{7p} A. Ramos, C. Arias, E. Fuenmayor and E. Contreras,  {\it Eur. Phys. J. C} {\bf 81}, 203 (2021).

\bibitem{LN3} H. Hern\'andez, D. Su\'arez--Urango and L. Nu\~nez, {\it Eur. Phys. J. C} {\bf 81}, 241 (2021).

\bibitem{efl} P. Le\'on, E. Fuenmayor and E. Contreras {\it Phys. Rev.  D} {\bf 104}, 044053 (2021).

\bibitem{LN2} D. Su\'arez--Urango, J. Ospino, H. Hern\'andez and L. Nu\~nez, {\it Eur . Phys .J. C} {\bf 802}, 176 (2022). 

\bibitem{45} T. Feroze and A. A. Siddiqui, {\it Gen. Relativ. Gravit.} {\bf 43}, 1035 (2011).

\bibitem{46} S. Mardan, A. Asif, and I. Noureen, {\it Eur . Phys .J. C} {\bf 134}, 1 (2019).

\bibitem{ILEP} D. Santana, E. Fuenmayor and E. Contreras, {\it Eur. Phys. J. C} {\bf 82}, 703 (2022).

\bibitem{JO} J. Ovalle, {\it Phys. Rev. D} {\bf 95}, 104019 (2017).

\bibitem{Mazur:2001fv}
P.~O.~Mazur and E.~Mottola,
Universe \textbf{9} (2023) no.2, 88
doi:10.3390/universe9020088
[arXiv:gr-qc/0109035 [gr-qc]].

\bibitem{Tolman} R. Tolman, {\it Phys. Rev.}  {\bf 35}, 875 (1930).

\bibitem{15} L. Herrera, A. Di Prisco, J. Hern\'andez-Pastora, and N. O. Santos, {\it Phys. Lett. A} {\bf 237}, 113 (1998).

\bibitem{hod08} L. Herrera, J. Ospino and A. Di Prisco, {\it Phys. Rev. D} {\bf 77}, 027502 (2008).

\bibitem{Cosenza} M. Cosenza, L. Herrera, M. Esculpi, L. Witten, {\it J. Math. Phys.} {\bf 22}, 118 (1981).

\bibitem{Herrera2001} L. Herrera, A. Di Prisco, J. Ospino, and E. Fuenmayor, {\it J. Math. Phys.} (N.Y.) {\bf 42}, 2129 (2001).

\bibitem{VC}  L. Herrera, {\it Phys. Rev. D} {\bf 97}, 044010 (2018).

\bibitem{bh} P. Bargue\~no, E. Fuenmayor and E. Contreras, {\it Annals Phys.} {\bf 443}, 169012 (2022).

\bibitem{durgapal} E. Contreras, E. Fuenmayor2 and G. Abellán2, {\it Eur. Phys. J. C.} {\bf 82}, 187 (2022).

\bibitem{zh} Ya. B. Zeldovich, Zh. Eksp. Teor. Fiz. {\bf 41}, 1609 (1969) [Sov. Phys. JETP  {\bf 14}, 1143 (1962)].

\bibitem{VC2}  C. Arias, E. Contreras,  E. Fuenmayor and A. Ramos, {\it Ann. Phys.} {\bf 436}, 168671 (2022).

\bibitem{karmarkar} K. R. Karmarkar, {\it Proc. Indian Acad. Sci. A} \textbf{27}, 56 (1948).



\end{thebibliography}
\end{document}